\documentclass[twocolumn,amsmath,amssymb,floatfix,showpacs,prb]{revtex4}

\usepackage{graphicx}
\usepackage{dcolumn}
\usepackage{bm}
\usepackage{color}
\usepackage{hyperref}

\graphicspath{{./}{./figs/}}

\begin{document}
\title{Transition voltage spectroscopy: a challenge for vacuum tunneling models at nanoscale}

\author{Ioan B\^aldea}
 \altaffiliation[Also at ]{National Institute for Lasers, Plasma, and Radiation Physics, 
ISS, POB MG-23, RO 077125, Bucharest, Romania} 
\author{Horst K\"oppel}%
\affiliation{%
Theoretische Chemie, Universit\"at Heidelberg, Im Neuenheimer Feld 229, 
D-69120 Heidelberg, Germany}

\begin{abstract}

Several recent studies on the transition voltage ($V_t$) of molecular and vacuum nano-junctions 
are based on calculations of the tunneling current through an energy barrier 
within the so-called Simmons model, which is an approximate WKB-type approach
developed for thin insulating films with infinite transverse extension. 
In this paper devoted to vacuum nano-junctions, 
we compare the Simmons results for $V_t$ 
with those obtained from the exact Schr\"odinger equation by exactly including 
the classical (non-retarded) charge image effects. The comparison reveals that 
the Simmons estimates for $V_t$ are completely unacceptable for nanogap sizes ($d$) 
at which image effects are important. 
The Simmons treatment drastically overestimates these effects, because it misses the
famous 1/2
factor related to the fact that the image interaction energy is a self energy.
The maximum of the Simmons curve $V_t$ vs.~$d$ turns out to be merely an artefact 
of an inappropriate approximation. Unlike the Simmons approach, 
the ``exact'' WKB method yields results, which qualitatively agree with the exact ones;
quantitative differences are important, demonstrating that the transmission 
prefactor has a significant impact on $V_t$.
Further, we show that a difference between the work functions of (say,) the left and right 
electrodes, which gives rise to a Volta intrinsic field, may be important for the 
ubiquitous asymmetry of the measured $I$-$V$-characteristics [$I(V) \neq -I(-V)$]
in general, and for the different $V_t$-values at positive and negative biases
reported in vacuum nano-junctions in particular. 
The weak dependence $V_t = V_t(d)$ found experimentally in vacuum nano-junctions contrasts 
to the pronounced dependence obtained by including the exact electrostatic 
image contribution into the exact Schr\"odinger equation. This demonstrates that 
not only the molecular transport, but also the transport through a vacuum nano-gap 
represents a nontrivial problem, which requires further refinements, 
e.~g, a realistic description of contacts' geometry, 
retardation effects due to the finite tunneling time, 
local phonons, surface plasmons or electron image states.
\end{abstract}
\pacs{
73.63.Rt, 
85.35.Gv  
85.65.+h,  
}
\keywords{molecular electronics, single-electron transistors, transition voltage spectroscopy, 
Fowler-Nordheim transition}
\maketitle

\section{Introduction}
\label{sec:introd}
In the continuous efforts for miniaturization, 
using single molecules as 
active components for future nanoelectronic devices appears at present
as the only conceivable alternative, which escapes the fundamental limitations 
of complementary metal-oxide semiconductor (CMOS) technologies. In 
molecular devices, electron transfer between the source ($S$) and drain ($D$) 
electrodes 
across the nanogap (width $d$) can occur via through-bond and through-space processes.
In these processes, electrons have to tunnel through an energy barrier 
given by 
the energy offset 
$\varepsilon_B = \min 
\left( \varepsilon_F - \varepsilon_{HOMO}, \varepsilon_{LUMO} - \varepsilon_F \right)$
of the
closest molecular orbital (HOMO or LUMO) from electrodes' Fermi level ($\varepsilon_F$)
and the metallic work function 
$W$,
respectively. Similar to electron transfer
in numerous chemical reactions, through-bond and through-space processes can compete,
because (i) in usual off-resonance situations, 
the Fermi level lies close to middle of the HOMO-LUMO gap
(charge neutrality), and 
$\varepsilon_{B}$ amounts several electronvolts,
i.~e., only slightly smaller than 
$W$
and (ii) charge image effects, 
which narrow, round, and lower the barrier, 
\cite{Sommerfeld:28,Sommerfeld:33} are more pronounced in vacuum than in molecules 
due to the different dielectric constants [$\kappa_r \equiv 1$ versus $\kappa_r \approx 2 - 3$, 
respectively, cf.~Eq.~(\ref{eq-phi-i}) below]. Concerns that the measured currents 
are not (or not only) due to the active molecules have plagued the field of molecular electronics
since its inception.\cite{aviram74} For a proper interpretation of transport experiments, 
through-bond and through-space mechanisms should be each correctly understood. 

Transition voltage spectroscopy (TVS) has been proposed recently as an appealing 
tool to deduce $\varepsilon_B$, a key parameter for molecular junctions.
\cite{Beebe:06,Choi:08,Beebe:08,Frisbie:10b} 
TVS 
relies upon simple intuitive considerations inspired by the barrier picture. 
The initial TVS conjecture was that
the minimum $V=V_t$ of the Fowler-Nordheim (FN) curves [$\log(I/V^2)$ versus $1/V$, 
obtained by recasting the source-drain $I$-$V$-characteristics] suffices to determine the 
molecular energy offset: $e V_t = \varepsilon_{B}$. This minimum was ascribed
to the point where the shape of the energy barrier 
tilted by the applied voltage $V$ 
($\varepsilon_{B} \to \varepsilon_{B} - e V x/d$, where $x$ specifies the position)
changes from trapezoidal to triangular.
Calculations of the tunneling current 
within the Simmons' model \cite{Simmons:63} 
challenged the validity of 
the barrier description for molecular junctions,\cite{Huisman:09} 
claiming that it yields a dependence 
of $V_t$ on $\varepsilon_{B}$ and $d$ disagreeing with experiments.
\cite{Beebe:06,Choi:08,Beebe:08,Choi:10,Frisbie:10b} 
Based on them, it was 
suggested that the dependence $V_t = V_t(d)$ \cite{Huisman:09} or 
the magnitude of $V_t$ \cite{Molen:11} 
can be used to discriminate between through-bond and vacuum tunneling.
Some aspects of the TVS in molecular junctions 
\cite{Huisman:09,Araidai:10,Thygesen:10b,Thygesen:11} and gated single-molecule
transistors \cite{Baldea:2010h} received theoretical consideration.

Letting alone the fact that the calculations of Ref.~ \citenum{Huisman:09}
used the Simmons' approximation \cite{Simmons:63}
(an issue to be addressed in detail in the present paper), it is worth emphasizing that 
those calculations disregarded an important fact. Namely, that in order to phenomenologically 
model the 
tunneling through a molecule, it is too simplistic that, irrespective of the molecule under consideration, 
electrons within the barrier (``molecule'') be generally characterized by the free electron mass 
$m$. Several studies drew attention on the fact that this is an illegitimate assumption,\cite{Rampi:01,Reed:04}
and one rather needs to employ an effective mass $m^\ast$, which is molecule specific
[with the present notations $m^\ast = m^\ast(d, \varepsilon_{B})$]. 
For the above reasons, the utilization of the barrier picture for molecular transport is 
only possible if the phenomenological parameters $\varepsilon_{B}$ and $m^\ast$ 
are determined by fitting experimental data or from companion ab initio calculations.

While modeling the active molecule as an energy barrier may be too
simplistic, and describing the electrons that tunnel 
in molecules (unless they possess delocalized $\pi$ electrons) 
within the effective mass approximation questionable,
the barrier picture appears as the most natural framework for
studying the transport in vacuum nanojunctions.
Still, it is only recently that a theoretical study becomes attractive,
because, in contrast to the immense number of studies 
on molecular junctions (a very incomplete list includes 
Refs.~\citenum{Beebe:06,Choi:08,Beebe:08,Chiu:08,Reed:09,Choi:10,Frisbie:10b,Reed:10,Bergren:10,Reed:11,Lennartz:11} 
and citations therein) experiments on 
electron tunneling through vacuum in nanojunctions have been carried out only very 
recently.\cite{Molen:11}

Among very numerous approaches to the electron transport by tunneling relying upon 
the barrier 
picture,\cite{FowlerNordheim:28,Nordheim:28,Sommerfeld:28,Sommerfeld:33,Holm:51,Bardeen:61,Harrison:61,Stratton:62,Hartman:64,Gundlach:66,Brinkman:70} 
an approach
due to Simmons \cite{Simmons:63} became increasingly popular for the interpretation
of important experimental findings in molecular-/nano-junctions.
\cite{Rampi:01,Frisbie:04,Reed:04,Beebe:06,Choi:08,Beebe:08,Reed:09,Choi:10,Frisbie:10b}
This popularity among experimentalists 
should certainly be related to the fact that, unlike in other (more accurate, see, e.~g., Ref.~\citenum{Gundlach:66})
treatments, Simmons gave approximate but simple analytic formulas,
which can be easily used for fitting experimental data. 
While primarily discussing and formulating the barrier approach to nanojunctions in general, 
we will also 
to draw attention on the fact that Simmons' formulas, aiming to describe 
thin insulating films, 
are not appropriate for molecular/nano-transport. An important issue to be emphasized 
is that 
certain predictions based on these formulas are artefacts, which preclude an adequate 
interpretation of valuable experimental findings.

The remaining part of this paper is organized as follows. 
In Section \ref{sec:review}, the Simmons approach will be reviewed, 
emphasizing its limitation. 
The confinement of electron motion in transverse directions, an aspect 
which is important for nanojunctions with atomic contacts but is not accounted for
within the Simmons approach, will be considered in Section \ref{sec:confinement}. 
Section \ref{sec:limitations} is devoted to the comparison of the results 
for $V_t$ obtained 
by solving the Schr\"odinger equation exactly with those of the 
``exact'' WKB and (``approximate'' WKB-type) Simmons approach, 
which clearly demonstrates that the last 
approach is inappropriate to account for image effects at sizes of interest
for nanotransport. In Section \ref{sec:W_L-W_R}, the case of electrodes with different
work functions will be analyzed. An important technical issue, the relationship 
between the apparent barrier height and the electrodes' work function, will be addressed
in Section \ref{sec:Delta-vs-W}. Based on the present exact results for $V_t$,
in Section \ref{sec:wkb} it will shown that the validity of the WKB approximation 
has been too optimistically assessed in previous studies.
The main results of the present study will be summarized in Section \ref{sec:conclusion},
where possible sources for the disagreement between the present theory and experiment
will be indicated.
\section{Reviewing the Simmons' approach}
\label{sec:review}
The theoretical framework of the single-particle tunneling through an energy barrier
was established long ago by Sommerfeld.\cite{Sommerfeld:33}
The current density due to the (elastic) tunneling along the $x$-direction 
across a sandwich consisting of an insulating film placed between two  
(source and drain) electrodes caused by a bias $V > 0$ between source 
(chosen below as the negative electrode)
and drain (positive electrode) with chemical potentials $\mu_{S} = 0$ and 
$\mu_{D} = - e V$ can be expressed as
\cite{Sommerfeld:33,Holm:51,Stratton:62,Simmons:63,Hartman:64,Gundlach:66,Brinkman:70}
\begin{eqnarray}
\displaystyle
\label{eq-J-3d}
& J & = 2 \frac{e}{h^3} \int_{\mu_{D} < E_{\mathbf{p}} < \mu_{S}} 
\frac{\partial E_{\mathbf p}}{\partial p_x}\, \mathcal{T}(p_{x}, V)\,{d^3\mathbf{p}} = \frac{4\pi m e}{h^3}\\
& & \times \left[ e V \hspace*{-1ex} \int_{-\varepsilon_F}^{-e V} \hspace*{-3ex} \mathcal{T}(E_{x}, V) d\,E_x 
\hspace*{-1ex} -
\int_{- e V}^{0}  \hspace*{-3ex} E_x \mathcal{T}(E_{x}, V) d\,E_x \right],
\label{eq-J-3d-0}
\end{eqnarray}
where $\mathcal{T}$ is the transmission coefficient.
To get Eq.~(\ref{eq-J-3d-0}), we assumed 
a parabolic conduction band dispersion 
$E_{\mathbf{p}} = \frac{1}{2m}\left(p^2_x + p^2_y + p^2_z\right) - \varepsilon_F$. 

To deduce the current, a very popular approach 
is to employ the WKB expression of the transmission
\begin{equation}
\displaystyle
\mathcal{T}_{WKB}(E_{x}, V) \equiv e^{- \frac{2 \sqrt{2 m}}{\hbar} 
\int_{\phi_{B}(x) > E_x} 
d\,x\,\sqrt{ \phi_{B}(x) - E_x} } . \label{eq-wkb-exact} 
\end{equation}

Like many others (e.~g,~Refs.~\citenum{Sommerfeld:28,Sommerfeld:33,Holm:51,Stratton:62,Hartman:64,Brinkman:70}), 
Simmons \cite{Simmons:63} 
did not use the above ``exact'' WKB approximation, but rather an approximate 
version of the ``exact'' WKB expression
\begin{eqnarray}
\mathcal{T}_{r}(E_{x}, V) & = &  
\exp\left( -A \sqrt{\overline{\varepsilon}_B - E_x}\right)  ,
\label{eq-wkb-approx} 
\end{eqnarray}
replacing the general position dependent barrier $\phi_B(x)$ by a rectangular 
barrier. 
In Eq.~(\ref{eq-wkb-approx}), the fact that the barrier $\phi_{B}(x)$ is not rectangular is accounted
for by considering an effective (average) rectangular barrier of height $\overline{\varepsilon}_B$
and a spatial extension $\Delta s$  that can be smaller than the 
distance between electrodes (embedded molecule/nanogap size) $d$ as well as
a correction factor $\gamma$. This factor enters the quantity 
$ A  \equiv  2 \gamma \Delta s (2 m)^{1/2}/\hbar$.
For certain insulating thin films, Simmons estimated $\gamma \simeq 1 $
within at most a few percents. A correction factor $\gamma \equiv 1$ has been implicitly
assumed in Refs.~\citenum{Huisman:09,Molen:11} although several earlier studies drew attention
on the fact that this assumption is inconsistent with experimental $I$-$V$-data
for molecular transport. 
Alternatively, in view of the above expression of $A$,
departures from a rectangular-shaped barrier \cite{Sommerfeld:33,Holm:51,Simmons:63}
can be considered by employing an effective mass $m^\ast \neq m$
($m \to m^\ast = m \gamma^2$).\cite{Rampi:01,Reed:04}

By using Eq.~(\ref{eq-wkb-approx}), Eq.~(\ref{eq-J-3d-0}) can be integrated exactly in 
closed analytical form. The resulting, rather lengthy expression was given previously 
by various authors
(e.~g., Refs.~\citenum{Sommerfeld:33,Holm:51}) including Simmons, \cite{Simmons:63} 
who only retained the leading terms in his numerical calculations.
The central result of the Simmons approach \cite{Simmons:63} is the following 
expression of the current density $J$
\begin{eqnarray}
\label{eq-j-simmons}
J / J_0 & = &
\overline{\varepsilon}_B 
\left\{1 + \mathcal{O}\left[ \left( A \overline{\varepsilon}_B^{1/2} \right)^{-1}\right]\right\}
e^{ - A \overline{\varepsilon}_B^{1/2}} - \left(\overline{\varepsilon}_B + e V \right) \nonumber \\
& \times & 
\left\{1 + \mathcal{O}\left[ \left( A \left(\overline{\varepsilon}_B + e V\right)^{1/2} \right)^{-1}\right]\right\}
e^{ - A \left( \overline{\varepsilon}_B + e V\right)^{1/2}} \nonumber \\
& + &  
\mathcal{O}\left[
e^{ - A \left(\overline{\varepsilon}_B + \varepsilon_F\right)^{1/2} }
\right] , 
\end{eqnarray}
where $J_0 \equiv  e/\hbar/ (2 \pi\gamma \Delta s)^2$.
Eq.~(\ref{eq-j-simmons}) applies if the Fermi level of (say,) the right electrode
lies above the bottom of the left electrode, 
$0 < e V < \varepsilon_F$; otherwise the second term in the RHS 
(``backward'' current) should be omitted. 
This $V$-range suffices for practical purposes, since it 
is much broader than the range of the biases, which the molecular junctions fabricated 
so far can withstand or those used in experiments on vacuum nanojunctions.\cite{Molen:11}
Simmons' leading order expression (\ref{eq-j-simmons}) was utilized in 
a series of more recent works 
\cite{Rampi:01,Frisbie:04,Reed:04,Beebe:06,Choi:08,Beebe:08,Reed:09,Choi:10,Frisbie:10b}
to interpret a series of valuable experimental findings. 
An important pragmatical advantage of the approaches based on  
Eqs.~(\ref{eq-wkb-approx}) and (\ref{eq-j-simmons}) 
is that it can also simply account for image effects, 
see Eq.~(\ref{eq-eB-eff}). \cite{Sommerfeld:33,Holm:51,Simmons:63}

By supposing identical electrodes (work functions $W_{S} = W_{D} = W$) first,
the total barrier $\phi_B(x)$ through which electrons have to tunnel can 
be expressed as \cite{Sommerfeld:33,Simmons:63b,Gundlach:66}
\begin{equation}
\label{eq-phiB}
\phi_{B}(x) = \varepsilon_B - e V x /d + \phi_{i}(x) .
\end{equation}
In addition to a bare (rectangular) energy barrier $\varepsilon_B$ 
($\varepsilon_B = W$ for vacuum tunneling),
the energy barrier comprises the contributions of the applied bias $V$ and 
of the charge images $\phi_{i}(x)$.\cite{Sommerfeld:33}

The effective barrier width 
$\Delta s \equiv s_2 - s_1 $ is determined by Simmons
from the barrier extension ($0 \leq s_1 < x < s_2 \leq d$) at the electrodes' 
Fermi energy without bias (notice that energies are measured throughout relative to 
the Fermi level)
\begin{equation}
\label{eq-s1s2}
\phi_B(x)\vert_{x=s_{1,2}} = 0 .
\end{equation}
The effective barrier height is expressed as
\begin{equation}
\label{eq-eB-eff}
\overline{\varepsilon}_B = 
\varepsilon_B - e V \frac{s_1 + s_2}{2 d} + \int_{s_1}^{s_2} \frac{d\,x}{\Delta s} \phi_{i}(x) .
\end{equation}
In the range of experimental interest ($e \vert V \vert < \varepsilon_B$)
and ignoring image effects ($\phi_i \equiv 0$), the barrier is trapezoidal,
$s_1 = 0$, $\Delta s = s_2 = d$, and $\overline{\varepsilon}_B = \varepsilon_B - eV/2$.

For infinite planar electrodes placed at $x=0$ and $x=d$, the 
electrostatic interaction energy $\phi_i(x)$ 
between an electron located at $x$ and its images can be expressed exactly 
\cite{Sommerfeld:33}
\begin{equation}
\label{eq-phi-i}
\phi_{i}(x) = 
\frac{e^2}{4 \kappa_r d} \left[ - 2 \psi(1) + \psi\left(\frac{x}{d}\right) + 
\psi\left(1 - \frac{x}{d}\right)\right] ,
\end{equation}
where $\psi$ is the digamma function. To better understand the later analysis, 
we include throughout a dielectric constant $\kappa_r$ although for vacuum 
($\kappa_r \equiv 1$) it is superfluous.
In order to work out Eqs.~(\ref{eq-s1s2}) and (\ref{eq-eB-eff}) in closed analytic
forms, 
various parabolic approximations of the RHS of Eq.~(\ref{eq-phi-i})
around its maximum at $x=d/2$,
$ \phi_{i}\left(\frac{d}{2}\right) =  - \frac{e^2}{\kappa_r d} \log 2 $
were employed in Refs.~\citenum{Sommerfeld:33,Holm:51,Stratton:62}.
Instead of Eq.~(\ref{eq-phi-i}), Simmons used the 
``approximate'' expression
\begin{equation}
\label{eq-phi-i-simmons-approx}
\phi_{i}^{S}(x) \approx - 1.15 \log 2 \, \frac{e^2}{2 \kappa_r} 
\left( \frac{1}{x} + \frac{1}{d - x} \right) ,
\end{equation}
which he deduced from his ``exact'' expression
\begin{equation}
\label{eq-phi-i-simmons-ex}
\phi_{i}^{S}(x)  = 
-\frac{e^2}{\kappa_r}
\left\{ \frac{1}{2 x} + \sum_{n=1}^{\infty} \left[\frac{n d}{(n d)^2 - x^2} 
- \frac{1}{n d} \right] \right\} .
\end{equation}
In fact, Simmons' ``exact'' expression (\ref{eq-phi-i-simmons-ex}) is incorrect. The series 
entering the RHS of Eq.~(\ref{eq-phi-i-simmons-ex}) can be summed out, and the result is 
twice the RHS of Eq.~(\ref{eq-phi-i}). An easy way to understand this wrong factor is to consider  
situations very close to one electrode, e.~g., $x\agt 0$; then, Eq.~(\ref{eq-phi-i-simmons-ex})
yields $\phi_i(x) \simeq  - e^2/(2 \kappa_r x)$. This is twice the result
well known from electrostatics textbooks;
the attractive force between a point charge $e$ placed at $x > 0$ and an infinite 
plane at origin is $F_x (x) = - e^2/(4 \kappa_r x^2)$, and the corresponding interaction energy is 
$\phi_{i}(x) = - \int_{\infty}^{x} 
d\,\xi F_x (\xi) =  - e^2/(4 \kappa_r x)$. Simmons' exact expression misses 
nothing but the ``famous'' 1/2 
factor related to the fact that the image interaction energy $\phi_i(x)$ is a self energy 
and not the charge-potential product.\cite{Weinberg:82}

Recently, Simmons' expression (\ref{eq-phi-i-simmons-approx}) has been taken over in 
studies on image effects on the transition voltage.\cite{Huisman:09,Molen:11} 
In those works, all the terms (i.~e., not only the leading ones written explicitly) 
in Eq.~(\ref{eq-j-simmons}) were used in numerical calculations. To obtain the results 
presented below referred to as the Simmons results, we have also used 
these full expressions, as previously done in Refs.~\citenum{Huisman:09,Molen:11}. 
The terms $\mathcal{O}(\ldots)$ which are not explicitly 
written in Eq.~(\ref{eq-j-simmons}) may yield ``corrections'' to $V_t$ 
up to $\sim 10$\%,\cite{Bergren:10} for smaller values of $d$ and $\overline{\varepsilon}_{B}$
($A \overline{\varepsilon}_{B}^{1/2} \sim 1$). Still, we emphasize that the inclusion of these 
terms is a priori questionable, as Eq.~(\ref{eq-j-simmons}) is basically the result of a
(simplified) WKB approximation, and a valid WKB treatment requires sufficiently 
large values of $A \overline{\varepsilon}_{B}^{1/2}$ (see Ref.~\citenum{Gundlach:69b}
and the discussion below).
 
As a curiosity, we note that in order to improve the agreement between the theory based on
the Simmons approach and the experimental data for vacuum tunneling, 
a correction factor $\zeta \sim 0.4 - 0.7$ multiplying 
the RHS of Eq.~(\ref{eq-phi-i-simmons-approx}) has been empirically 
introduced in Ref.~\citenum{Molen:11}. It would be tempting to identify 
$\zeta$ with the aforementioned factor 1/2 missing in Eqs.~(\ref{eq-phi-i-simmons-ex}) 
and (\ref{eq-phi-i-simmons-approx}).
Unfortunately, the remedy of the Simmons results is not so simple, 
as it will be shown below.
\begin{figure}[h!]
$ $\\[6ex]
\centerline{\hspace*{-0ex}\includegraphics[width=0.4\textwidth,angle=0]{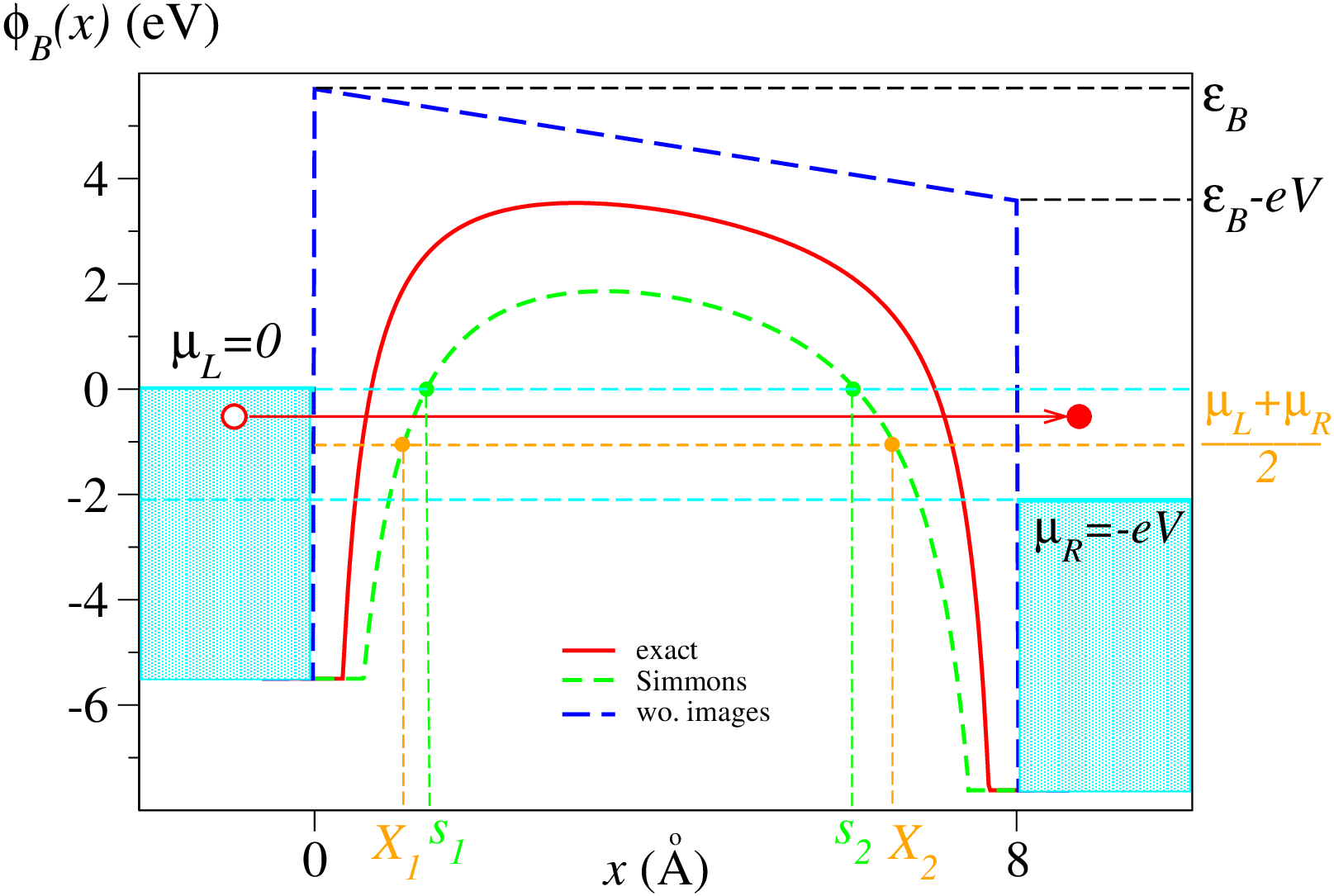}}
\caption{Energy barrier $\phi_B(x)$ 
for $\varepsilon_B \equiv W = 5.7$\,eV, $d=8$\,{\AA}, and $\kappa_r = 1$
computed using the exact contribution of charge images, Eq.~(\ref{eq-phi-i}), 
and using the Simmons' ``approximation'', Eq.~(\ref{eq-phi-i-simmons-approx}),
at $V\equiv V_t = 2.12$\,V. Notice that the height and width of the Simmons barrier are 
substantially smaller than the exact ones, and therefore image effects are 
drastically overestimated by the Simmons model.}
\label{fig:phi-B}
\end{figure}

Let us summarize the basic features of the Simmons' approach:

(i) Electrons tunnel across an energy barrier $\phi_{B}(x)$ in $x$-direction and move freely
in transverse ($y$, $z$)-directions, cf.~Eq.~(\ref{eq-J-3d}). 

(ii) The transmission coefficient across the barrier is computed within an approximation 
\emph{more} restrictive than the WKB-approximation, which assumes that the barrier profile 
$\phi_{B}(x)$ can be replaced by a rectangular effective barrier 
\begin{equation}
\label{eq-average}
\int \frac{d\,x}{\Delta s} \sqrt{ \phi_B(x) - E_x} \approx \gamma \sqrt{\overline{\varepsilon}_{B} - E_x}
\approx \sqrt{\overline{\varepsilon}_{B} - E_x} ,
\end{equation}
where the ($E_x$- and $V$-independent)
correction factor can be taken $\gamma \simeq 1$. 
This amounts to consider that, irrespective of their energy ($\mu_{D} < E_x < \mu_{S}$), all electrons that 
contribute to the current tunnel across a barrier of the same width $\Delta s = s_2 - s_1$
and height $\overline{\varepsilon}_{B}$, which is the barrier ``seen'' by the electrons 
at the Fermi level of the unbiased junction ($E_x = 0$), cf.~Eqs.~(\ref{eq-s1s2}) and (\ref{eq-eB-eff}), 
and Fig.~\ref{fig:phi-B}. 

(iii) The contribution of charge images to the energy barrier is estimated from 
Eq.~(\ref{eq-phi-i-simmons-approx}). This overestimates the 
image effects by a factor $\sim 2$.\cite{why-1.15}

To demonstrate the fact that the Simmons approach is inadequate for nanojunctions, we will compare 
below the results of this approach both with the results of the ``exact'' WKB 
approximation [i.~e., by using Eq.~(\ref{eq-wkb-exact}) and not 
Eq.~(\ref{eq-wkb-approx})]
as well as with the exact results obtained by solving the Schr\"odinger equation 
with the exact barrier potential, Eq.~(\ref{eq-phiB}). 
The exact transmission $\mathcal{T}(E_x, V)$ can be easily 
obtained by imposing the overall
continuity of the wave function and its derivative. 
Inside the junction ($0 < x < d$), 
the wave function $\Psi(x)$ obeys the Schr\"odinger equation
\begin{equation}
\label{eq-psi-M} 
\left[ 
- \frac{\hbar^2}{2 m} \frac{\partial^2}{\partial x^2} + 
\phi_{B}(x) - E_x \right] \Psi(x) = 0 .
\end{equation}
The exact Schr\"odinger 
equation written above can easily solved analytically in the absence of images 
($\phi_i \equiv 0$).\cite{Gundlach:66}
The exact wave function $\Psi(x)$ for this case can be
expressed as a linear combination of the Airy functions $Ai(z)$ and $Bi(z)$,
where $z \equiv \zeta - x/\lambda$ and
\begin{eqnarray*}
\displaystyle
\zeta & \equiv & \left[ \frac{2 m^{\ast}}{\hbar^2} 
\left(\frac{d}{e V}\right)^2\right]^{1/3}\mbox{\hspace*{-3ex}}
\left( \varepsilon_B - E_x\right); \mbox{ }
\frac{1}{\lambda} \equiv \left( \frac{2 m^{\ast}}{\hbar^2} \frac{e V}{d} \right)^{1/3} .
\end{eqnarray*}
The transmission is given by 
$\mathcal{T}(E_x, V)  =   \mathcal{N}(E_x, V) / \mathcal{D}(E_x, V)$,
where \cite{Gundlach:66}
\begin{eqnarray*}
\displaystyle
\mathcal{N} & = & 4 \frac{k_{S} k_{D}}{d^2}
\left( A_{D} B_{D}^{\prime} - A_{D}^{\prime} B_{D} \right)^2 ,  
\\
\mathcal{D} & = &
\left[
\frac{1}{d^2} \left( A_{S}^{\prime} B_{D}^{\prime} - A_{D}^{\prime} B_{S}^{\prime} \right)
+ k_{S} k_{D} \left( A_{S} B_{D} - A_{D} B_{S} \right)
\right]^2  \nonumber \\
& + & \left[ 
\frac{k_{S}}{d} \left( A_{S} B_{D}^{\prime} - A_{D}^{\prime} B_{S} \right)
+ \frac{k_{D}}{d} \left( A_{D} B_{S}^{\prime} - A_{S}^{\prime} B_{D} \right)
\right]^2 . \nonumber
\end{eqnarray*}
Above, the obvious dependence on $E_x$ and $V$ has been omitted for brevity,
the subscripts $S$ and $D$ correspond $x = 0$ and $x = d$, respectively,
and the prime stands for the derivative of the Airy functions
with respect to $z$, e.~g., $A_{D} \equiv Ai(\zeta - d/\lambda)$ and
$B_{S}^{\prime} \equiv \left[d\,Bi(z)/d\,z\right]_{z = \zeta}$, 
$k_{S} \equiv \sqrt{2 m (\varepsilon_F + E_x)}/\hbar$, and $k_{D} \equiv \sqrt{2 m (\varepsilon_F + E_x + eV)}/\hbar$.

In the  presence of image effects,
we have performed exact numerical calculations 
for the potential given by Eqs.~(\ref{eq-phiB}) and (\ref{eq-phi-i}) or 
(\ref{eq-phi-i-simmons-approx}). To this aim,
we used a sufficiently large number $N$ of small pieces (until reaching the convergence)
$0 \equiv x_0 \leq x_1 \leq \ldots \leq x_N \equiv d$ and considered 
piecewise constant potentials 
$\phi_{B,n} = \int_{x_{n-1}}^{x_n}\phi_{B}(\xi)/(x_{n} - x_{n-1})$.
\section{Lateral confinement}
\label{sec:confinement}
Eq.~(\ref{eq-j-simmons}) represented
the starting point for interpreting many valuable experimental results for molecular junctions. 
Therefore, it is noteworthy that, in fact, the description based on Eq.~(\ref{eq-J-3d}), 
from which Eq.~(\ref{eq-j-simmons}) is
deduced,
is appropriate neither for molecular junctions
nor for vacuum nanojunctions with atomic contacts, 
because in those cases the electron motion transverse to the 
junction is confined. Rather than being continuous variables
($0 < E_{y,z} < \varepsilon_F$ at zero temperature), the transverse energies 
are quantized, and only the lowest-energy transverse channel contributes to 
electric conduction. Therefore, to describe electric transport in such junctions,
the integration over the transverse ($y$, $z$) directions should be omitted, 
and the counterpart of 
Eqs.~(\ref{eq-J-3d}) and (\ref{eq-J-3d-0}) is a current expressed by
\begin{equation}
I = 2\frac{e}{h} \int_{- e V}^{0} \mathcal{T}(E_{x}, V) d\,E_x .
\label{eq-I}
\end{equation}
It is worth emphasizing that Eq.~(\ref{eq-I}) qualitatively differs from Eq.~(\ref{eq-J-3d-0}).
To understand that Eq.~(\ref{eq-J-3d-0}) is basically inappropriate
for nanotransport, let us consider the linear conductance
\begin{equation}
\label{eq-g-3d}
G \propto \lim_{V\to 0} \frac{d\,J(V)}{d\,V} 
\propto
\int_{0}^{\varepsilon_F} \mathcal{T}(E_x, V=0) d\,E_x .
\end{equation}
Eq.~(\ref{eq-g-3d}) is inconsistent with Landauer's fundamental statement 
for the transport at nanoscale (``conductance is transmission'',
$G \propto \mathcal{T}(\varepsilon_F, V=0)$). The
electrons contributing to Eq.~(\ref{eq-g-3d}) can have a kinetic energy 
$E_x$ of motion across the junction spanning the whole conduction band 
($0 \leq E_x \leq \varepsilon_F$); $E_x$ is not restricted to $E_x\simeq \varepsilon_F$.
This is the consequence of the fact that the lateral ($y,z$) 
confinement is missing in Eq.~(\ref{eq-J-3d-0}). The states contributing to
linear response in Eq.~(\ref{eq-J-3d}) have a total energy 
$E_{\mathbf{p}} \equiv E_x + E_y + E_z \simeq \varepsilon_F$,
but the transverse kinetic energies are continuous variables in the range 
$0 < E_{y,z} < \varepsilon_F$. To conclude,  Eq.~(\ref{eq-J-3d-0}) 
may be appropriate for
traditional electronics envisaged by Simmons, but not for nanojunctions with atomic contacts,
wherein the lateral confinement is essential and Eq.~(\ref{eq-I}) should be used.

Mathematically speaking, the only difference between the situation without lateral confinement 
discussed in Section \ref{sec:limitations} and that with lateral confinement is the manner 
of carrying out the energy integration, namely, Eq.~(\ref{eq-J-3d-0}) versus Eq.~(\ref{eq-I}).
Whether computed exactly by solving the Schr\"odinger equation (\ref{eq-psi-M}) 
or approximately via Eq.~(\ref{eq-wkb-exact}) or Eq.~(\ref{eq-wkb-approx}) 
the transmission coefficient is given by the same formula. This is the main reason
why the shortcomings of the Simmons results presented below 
are essentially the same, irrespective
whether the one-dimensional or three-dimensional description, underlying  
Eqs.~(\ref{eq-I}) or (\ref{eq-J-3d-0}), respectively is employed.
\section{Simmons' results versus WKB and exact results}
\label{sec:limitations}
In the cases typical for thin dielectric/semiconducting films, 
with (practically) infinite transverse extension and widths $d\sim 20 - 50$\,{\AA}, 
large dielectric constants $\kappa_r \sim 10$ 
[cf.~Eqs.~(\ref{eq-phi-i}) and (\ref{eq-phi-i-simmons-approx})],
and voltages that are not too high
analyzed by Simmons,\cite{Simmons:63} the approximations underlying Eq.~(\ref{eq-j-simmons})
may be reasonable.   

However, we will show that the above approximations are not 
justified for molecular/nano-junctions. Although we are going to present only results 
obtained by accounting for the lateral confinement discussed in 
Section~\ref{sec:confinement}, we note that there are only insignificant quantitative 
difference between the exact, WKB, and Simmons results 
also when the lateral confinement is ignored. To emphasize again,
the shortcomings of the Simmons approach 
discussed in this paper are not specific for the cases where the lateral confinement is important.

To obtain the numerical results presented below, we fixed the Fermi energy 
to the value $\varepsilon_F = 5.5$\,eV specific for gold, employed experimental
work function values $W_{S,D} = 3.2 - 5.7$\,eV (see also Section \ref{sec:Delta-vs-W} below),
and considered nanogap sizes $d$ relevant for experiments.\cite{Molen:11} 

Let us first examine the case of a relatively 
wide ($d=8$\,{\AA}) vacuum junction ($\kappa_r \equiv 1 $) with electrodes
characterized by identical work functions $\varepsilon_B = W_{S} = W_{D} = 5.7$\,eV 
(i.~e., the highest experimental value, which is the most favorable for a valid WKB treatment).
The consideration of vacuum nanojunctions obviates an important conceptual issue
for molecular junctions.
In the latter, in view of the finite tunneling time ($\tau$) spent by electrons within the barrier (``molecule''),
it is unclear whether the static dielectric constant $\kappa_r(\omega)\vert_{\omega = 0}$ 
or the dynamical dielectric $\kappa_r(\omega)\vert_{\omega \sim 1/\tau}$ should be used 
in Eq.~(\ref{eq-phi-i}).

\begin{figure}[h!]
$ $\\[6ex]
\centerline{\hspace*{-0ex}\includegraphics[width=0.4\textwidth,angle=0]{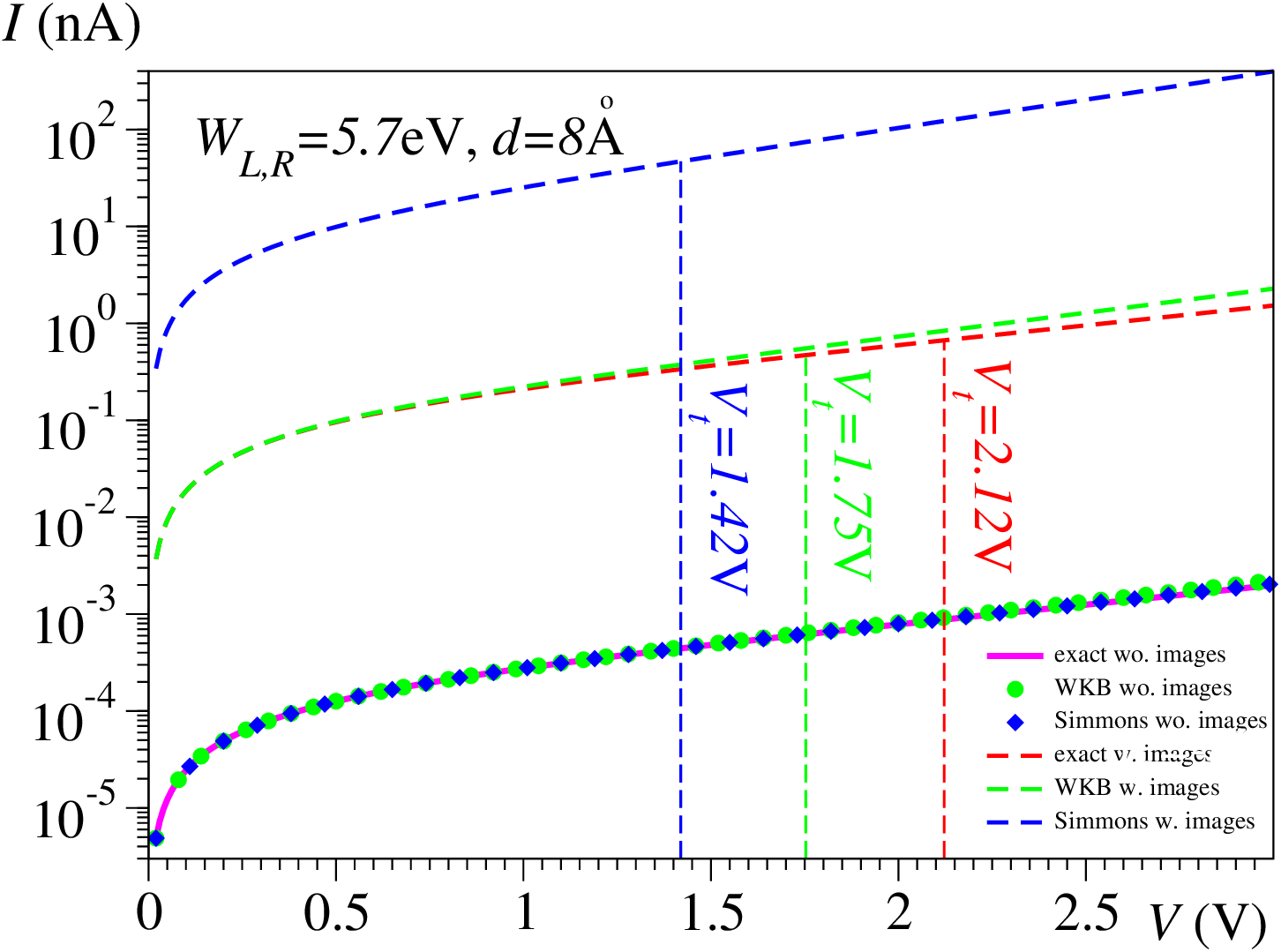}}
\caption{$I$-$V$-characteristics without and with images. 
Notice that the curve deduced from full quantum-mechanical calculations (label ``exact'')
without and with images has been multiplied with a factor 0.250 and 0.503, respectively, 
which represents the deviation 
of the ohmic conductance  of the ``exact'' WKB value from the ``exact'' conductance.}
\label{fig:iv-wo-w-image}
\end{figure}

\begin{figure}[h!]
$ $\\[6ex]
\centerline{\hspace*{-0ex}\includegraphics[width=0.4\textwidth,angle=0]{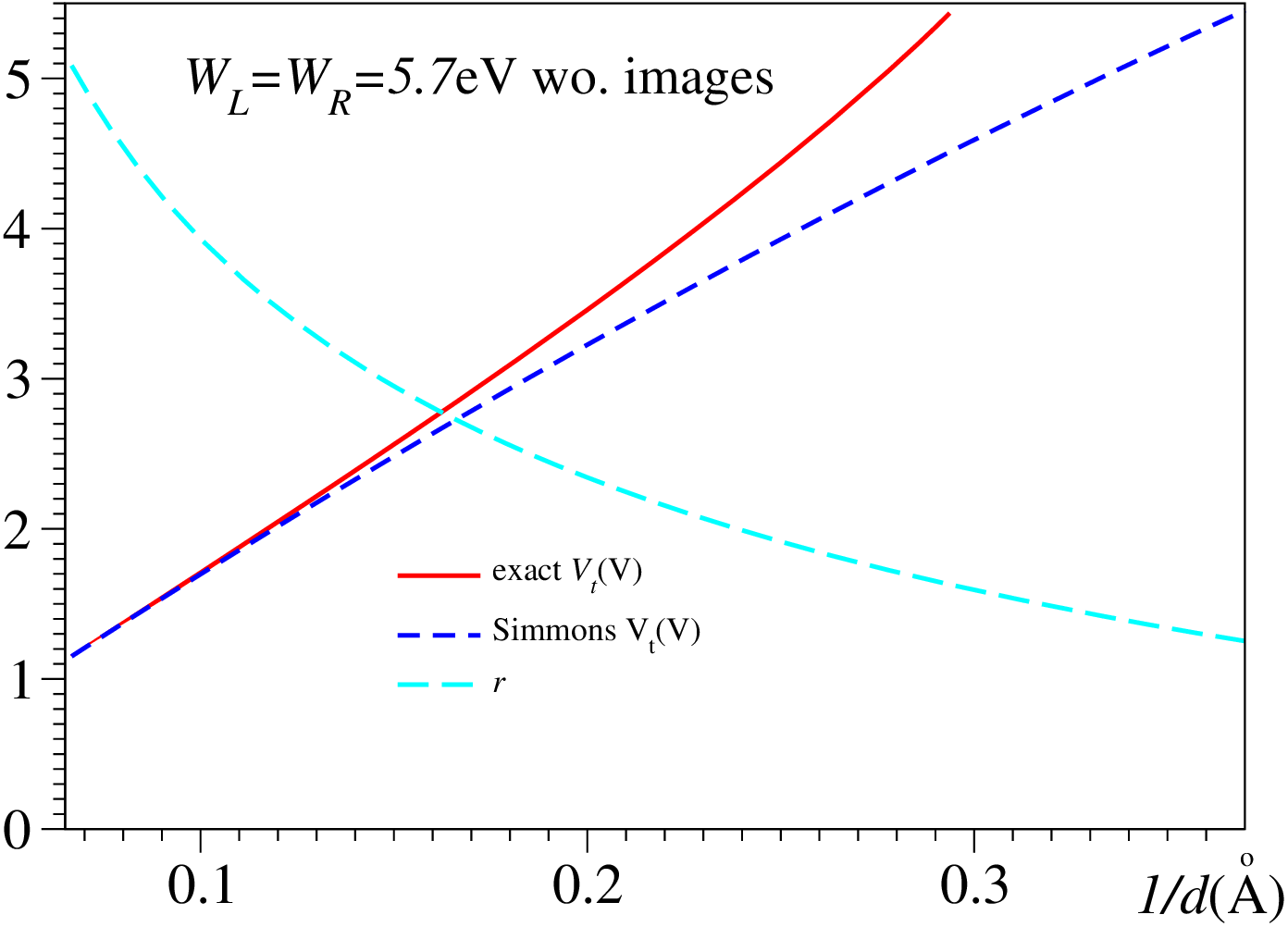}}
$ $\\[6ex]
\centerline{\hspace*{-0ex}\includegraphics[width=0.4\textwidth,angle=0]{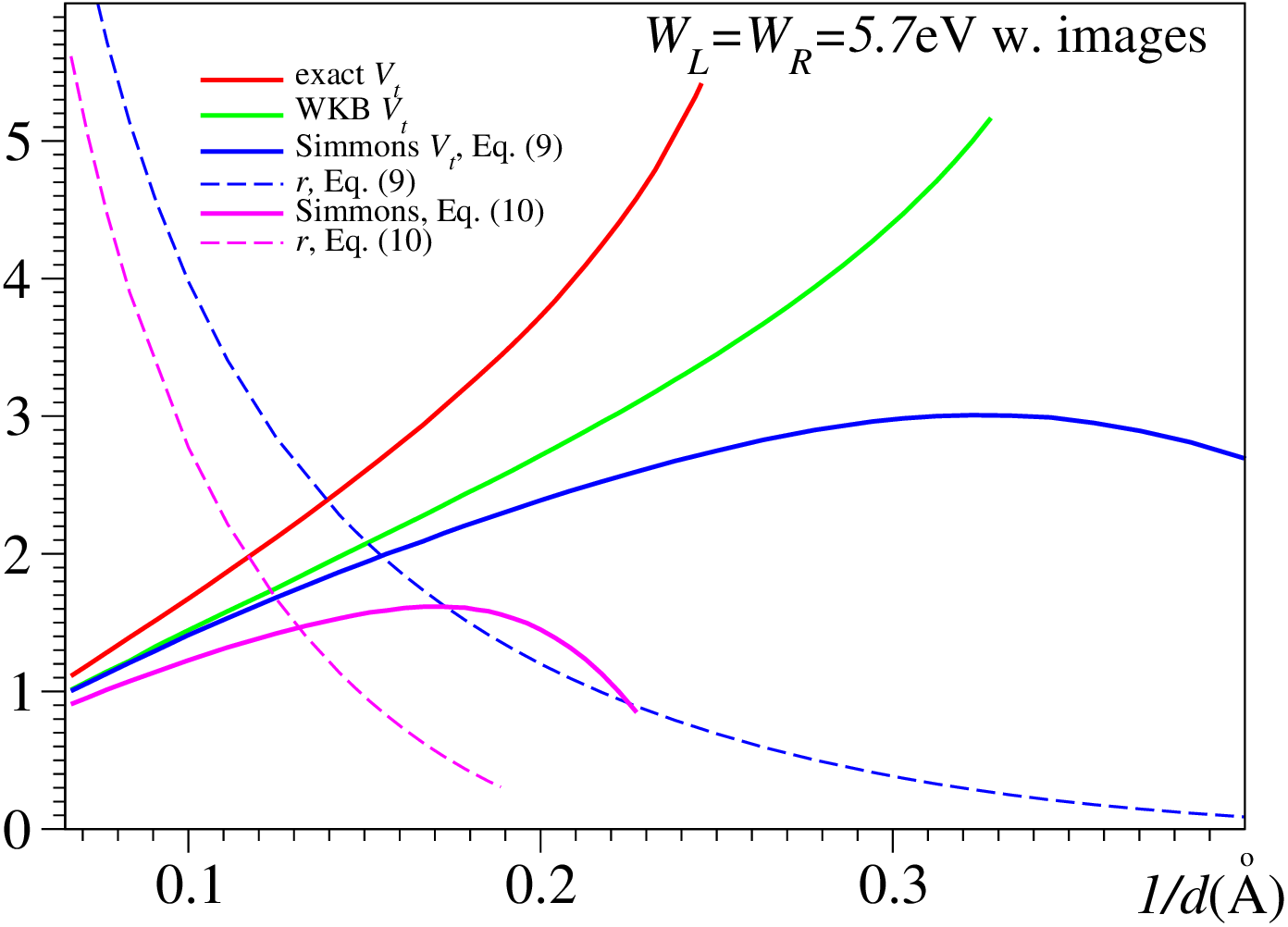}}
\caption{Exact, WKB, and Simmons results on the transition voltage for 
vacuum tunneling (a) without and (b) with charge images. In the latter case, results obtained 
within the Simmons approach are shown both with the incorrect and with the correct image forces
[Eqs.~(\ref{eq-phi-i-simmons-approx}) and (\ref{eq-phi-i}), respectively].
The curves for $r \equiv A {\overline{\varepsilon}_{B}}^{1/2}/4$ reveal 
that the condition $r > 1$ of Ref.~\citenum{Gundlach:69b} for a valid WKB approximation is too optimistic.}
\label{fig:vt-gundlach}
\end{figure}

Results for the vacuum nanogap specified above are presented in Fig.~\ref{fig:iv-wo-w-image}. 
The lower group of three curves
(which can hardly be distinguished among themselves
within the drawing accuracy) represent exact, WKB and Simmons results 
obtained by ignoring image effects.
For the bias range of experimental 
interest ($ V \alt 3$\,V),\cite{Molen:11} ignoring image effects,  
the Simmons approximation agrees well with the ``exact'' WKB approximation.
Concerning the latter, 
it basically deviates from the exact quantum-mechanical calculations by 
a \emph{constant} factor $I_{exact} \simeq 0.250 I_{WKB}$. 
Although this constant factor precludes a quantitative
analysis of experimental $I$-$V$-characteristics, it does not 
notably affect the transition voltage ($V_t = 2.13$\,V 
for the lower group of curves without images in Fig.~\ref{fig:iv-wo-w-image}), 
deduced from the minimum of $\log (I/V^2)$($ = \log I - 2 \log V$). 
For the highest work function 
$W_{S} = W_{D} = 5.7$\,eV deduced in experiments,\cite{Molen:11} 
a difference $\delta V_t \simeq 0.1$\,V (comparable to the 
experimental inaccuracy \cite{Beebe:06,Reed:09,Lennartz:11}) 
from the approximate estimates and the exact value 
sets the smallest nanogap size $d_{min} \simeq 8$\,{\AA} for 
reliable WKB and Simmons approximations for this case.
Briefly, the Simmons (and WKB) approximation yields reasonable estimates of the transition voltage $V_t$ 
for barriers sufficiently high and wide \emph{if} image effects were negligible.
However, this situation deteriorates at smaller sizes (see Fig.~\ref{fig:vt-gundlach}a)
and heights even without charge images.
\begin{figure}[h!]
$ $\\[6ex]
\centerline{\hspace*{-0ex}\includegraphics[width=0.4\textwidth,angle=0]{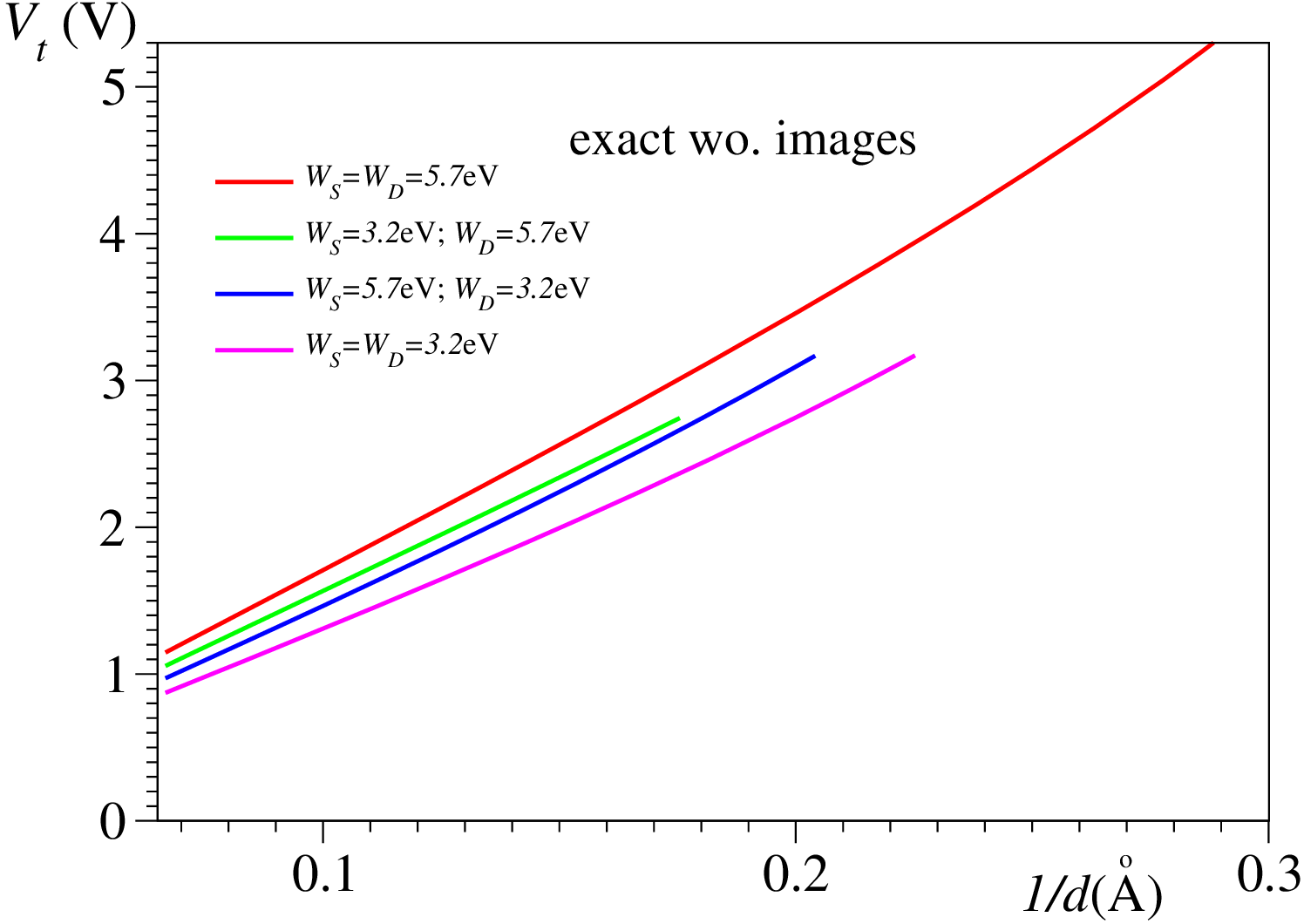}}
$ $\\[6ex]
\centerline{\hspace*{-0ex}\includegraphics[width=0.4\textwidth,angle=0]{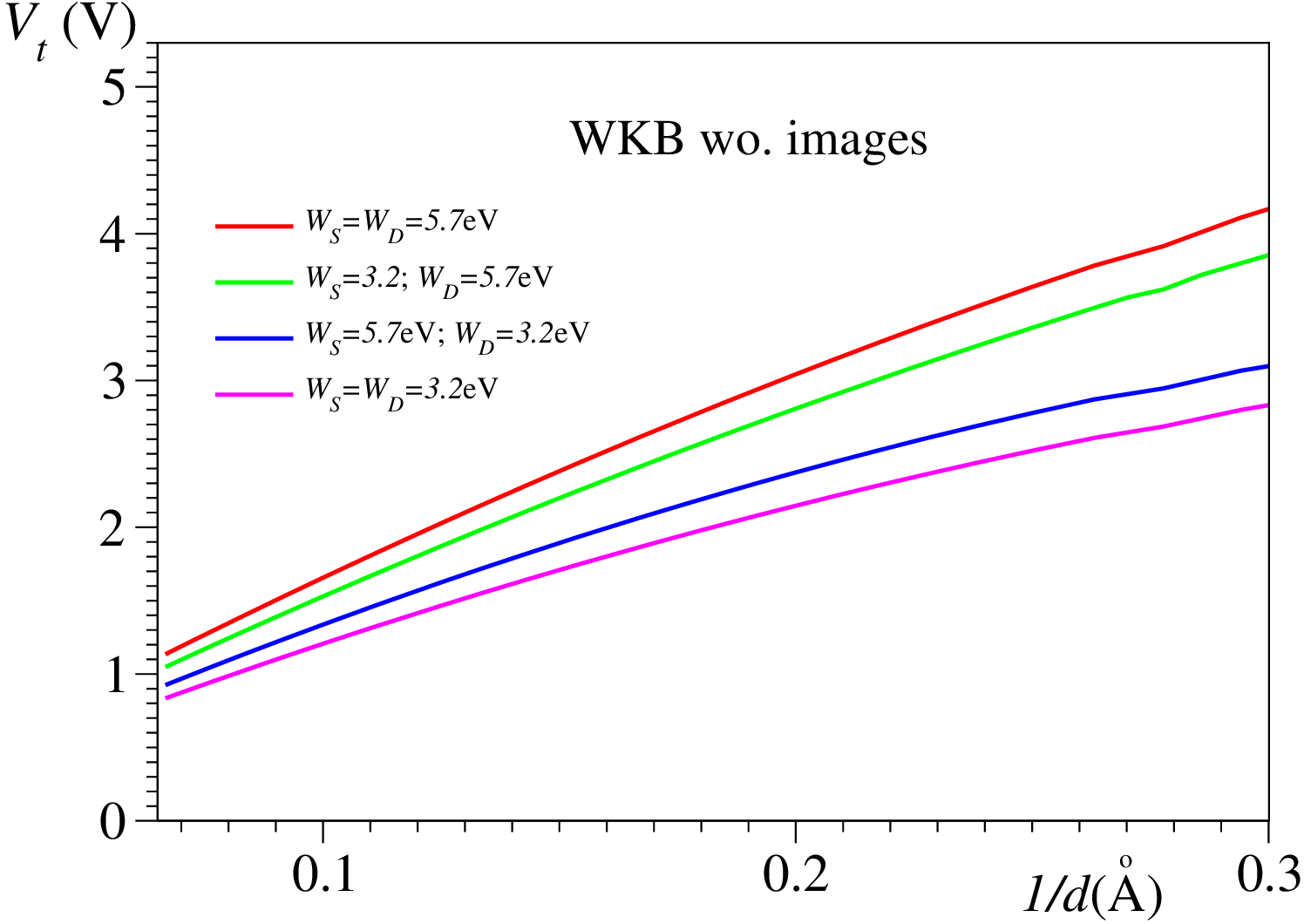}}
$ $\\[6ex]
\centerline{\hspace*{-0ex}\includegraphics[width=0.4\textwidth,angle=0]{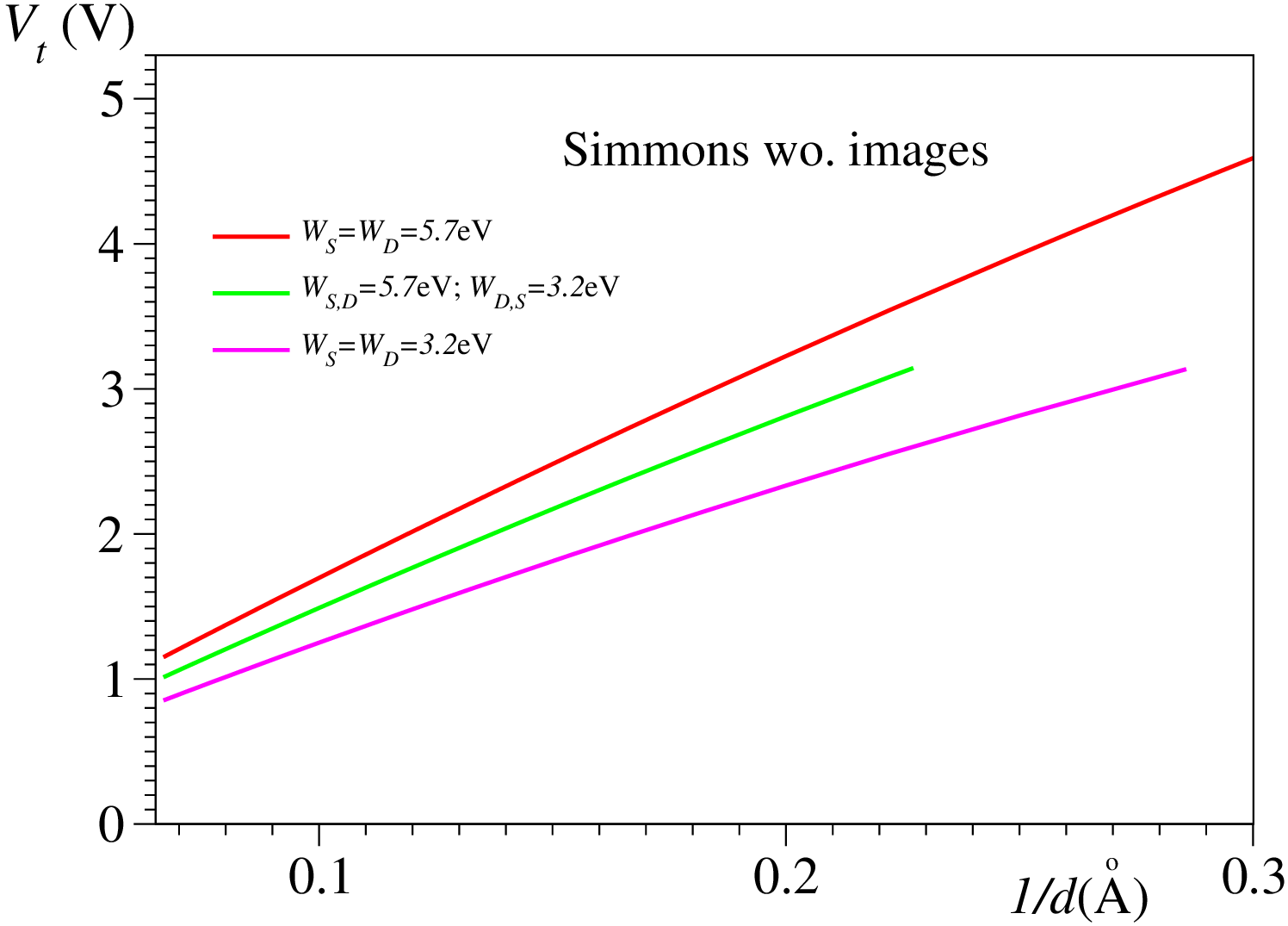}}
\caption{Transition voltage for vacuum tunneling without images computed exactly, and within the WKB and Simmons methods for electrodes with equal and different work functions $W_{S,D}$ specified in the legend.
Without image effects, the Simmons curves for ($W_{S} = 5.7$\,eV, $W_{D}=3.2$\,eV)
and ($W_{S} = 3.2$\,eV, $W_{D}=5.7$\,eV) coincide. 
Notice that, with our choice, the source/drain ($S$/$D$) is the negative/positive electrode
and $V_t$-values are always positive.}
\label{fig:vt-exact-wkb-huisman-n}
\end{figure}

The situation drastically changes when image effects are considered.
Because they rounds off the barrier corners, diminish and narrows the barrier hill, 
\cite{Sommerfeld:33} image effects increase the tunneling current. Image effects are most pronounced in
vacuum (lowest $\kappa_r = 1$).
For the case of Fig.~\ref{fig:iv-wo-w-image}, the exact current is enhanced by three orders of magnitude. 
Due to the incorrect factor 
in Eq.~(\ref{eq-phi-i-simmons-ex})
image effects are drastically exaggerated by the Simmons approach. 
The example of Fig.~\ref{fig:iv-wo-w-image} demonstrates that not only the Simmons 
results, but also the WKB values
significantly deviate 
from the exact ones. They become unacceptable at the relatively large 
barrier widths and heights
at which they would be reasonable without image effects.
This holds true not only for the magnitudes of the currents, but also for
the values estimated for the transition voltage. This is illustrated 
by the 
exact, WKB and Simmons results for the transition voltage in the presence of charge images,
which are presented in Figs.~\ref{fig:vt-gundlach}b
and \ref{fig:vt-exact-wkb-huisman-n}.
As visible there, Simmons' estimates for $V_t$ are acceptable only for large sizes ($d > 20$\,{\AA})
where image effects are negligible altogether.

Still, this is not the whole issue concerning the validity of the Simmons model
in cases where image effects are significant.
Most importantly, 
the Simmons approach even yields a \emph{qualitatively} incorrect prediction. This fact becomes
clear by inspecting the curves depicted in Fig.~\ref{fig:vt-gundlach}b, 
and by comparing the panels of Fig.~\ref{fig:vt-exact-wkb-huisman-i} among themselves.
As visible there, the Simmons curves are qualitatively incorrect;
they exhibit a maximum, which the exact curves do not display. The trend toward smaller sizes 
is just opposite: while, roughly, $V_t$ is inversely proportional to $d$ at large sizes
in both cases, at low sizes the exact $V_t$-values increase superlinearly with $1/d$
whereas the Simmons estimates decrease with $1/d$.
Simmons curves exhibiting a maximum 
have been previously shown in Refs.~\citenum{Huisman:09,Molen:11}, 
where the three-dimensional description [Eq.~(\ref{eq-J-3d-0})] was
implicitly assumed.
The present comparison with the exact curves demonstrates that this maximum is 
nothing but an artefact of the Simmons approximation, and this applies 
irrespective whether a lateral confinement is accounted for or not.
The occurrence of this Simmons maximum for curves of $V_t(d)$ is an artefact 
of the WKB rectangular-barrier approximation [approximation (ii) of Section \ref{sec:review}],
while the fact that this maximum is located at such small sizes 
is the consequence of the incorrect factor of 
Eqs.~(\ref{eq-phi-i-simmons-ex}) and (\ref{eq-phi-i-simmons-approx}) 
(as noted under (iii) in Section \ref{sec:review}).
Computations based on the approximate transmission given by Eq.~(\ref{eq-wkb-approx})
using the image potential energy of Eq.~(\ref{eq-phi-i}) instead of
Eq.~(\ref{eq-phi-i-simmons-approx}) still yield a maximum in $V_t(d)$ albeit located 
at substantially smaller sizes. This fact is illustrated by Figs.~\ref{fig:vt-gundlach}b,
and \ref{fig:vt-exact-wkb-huisman-i}c,d.
\section{Asymmetry driven by different electrodes' work functions}
\label{sec:W_L-W_R}
So far, we have considered electrodes characterized by identical work functions ($W_{S} = W_{D}$).
This is a reasonable assumption for vacuum tunneling between planar crystalline electrodes with macroscopic 
transverse extensions consisting of the same metal (even) if their separation ($d$)
falls in the nanometer range. However, this is an unlikely situation in molecular/nano-electronics
with atomic contacts.
Even for electrodes of the same chemical nature (e.~g., gold), it is unlikely 
that the local crystal orientation at contacts, which is hard to control experimentally,  
is identical. This general feature has been confirmed by the experimental data 
from the vacuum tunneling junctions fabricated in Ref.~\citenum{Molen:11}, for
which values $W \sim 3.2 - 5.7$\,eV have been deduced 
(see also Section \ref{sec:Delta-vs-W} below). 
This fact strongly supports the idea
that not only the work functions characterizing different junctions are different,
but also the values characterizing the (say,) left and right electrodes of a 
\emph{given} 
junction can be different. This effect seems to be not 
properly considered in recent studies on nanoelectronic devices with atomic contacts,
but may be quite relevant, as already pointed out by Sommerfeld
[see Ch.~20(b) and Fig.~36 of Ref.~\citenum{Sommerfeld:33}].

In the case of electrodes with different work functions 
($W_{S} \neq W_{D}$), a nonvanishing Volta potential difference arises, 
which yields an intrinsic (Volta) field $(W_{D} - W_{S})/(e d)$
that favors the motion of electrons from the metal with the higher work function 
to that with the lower work function.\cite{Sommerfeld:33}
In such situations, the above Eq.~(\ref{eq-phiB}) should be generalized to express
the total barrier $\phi_B(x)$ to account for the Volta field as follows 
\cite{Sommerfeld:33,Simmons:63b,Gundlach:66}
\begin{equation}
\label{eq-phiB-different-W}
\phi_{B}(x) = \varepsilon_B + \left(W_{D} - W_{S} - e V \right) x /d + \phi_{i}(x) .
\end{equation}
It is worth emphasizing that the Volta field is significant: the above example
reveals that values of $W_{D} - W_{S}$ can be a few electronvolts, such that the Volta 
field can be even larger than the applied field, since molecular/nano-junctions can
hardly withstand voltages larger than $V \sim 2 - 3$\,V.\cite{Reed:09,Molen:11,integration-limits}

More or less asymmetric experimental $I$-$V$-characteristics for positive and negative 
biases [i.~e., $I(V) \neq -I(-V)$] 
are ubiquitous in the transport at nanoscale.
This is also the case of the vacuum nanojunctions of Ref.~\citenum{Molen:11}.
A consequence of this
symmetry breaking is the fact that the transition voltages for positive and negative 
biases $V$ are different, a fact clearly visible in Fig.~1 of Ref.~\citenum{Molen:11}.

\begin{figure}[htb]
$ $\\[6ex]
\centerline{\hspace*{-0ex}\includegraphics[width=0.35\textwidth,angle=0]{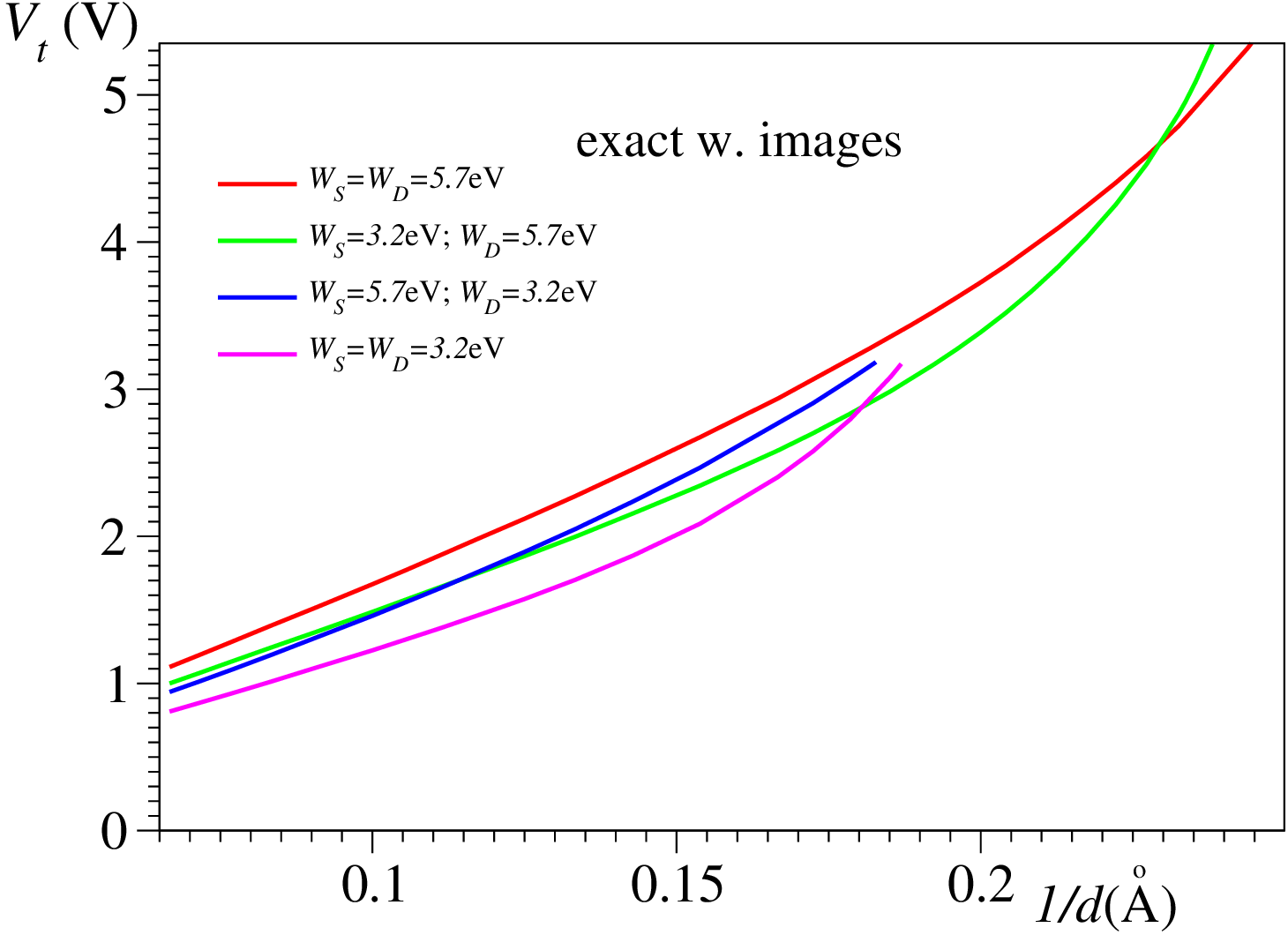}}
$ $\\[6ex]
\centerline{\hspace*{-0ex}\includegraphics[width=0.35\textwidth,angle=0]{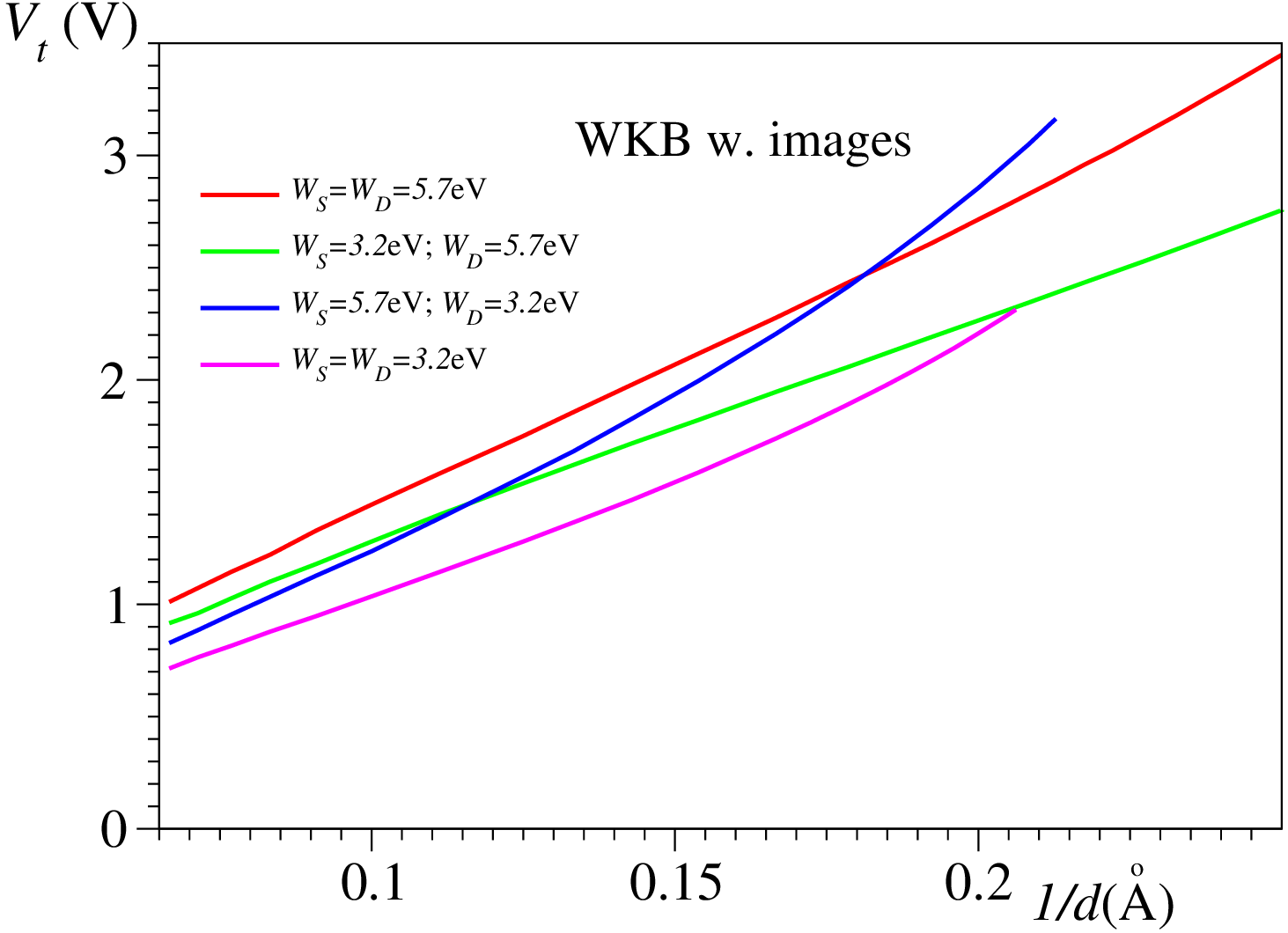}}
$ $\\[6ex]
\centerline{\hspace*{-0ex}\includegraphics[width=0.35\textwidth,angle=0]{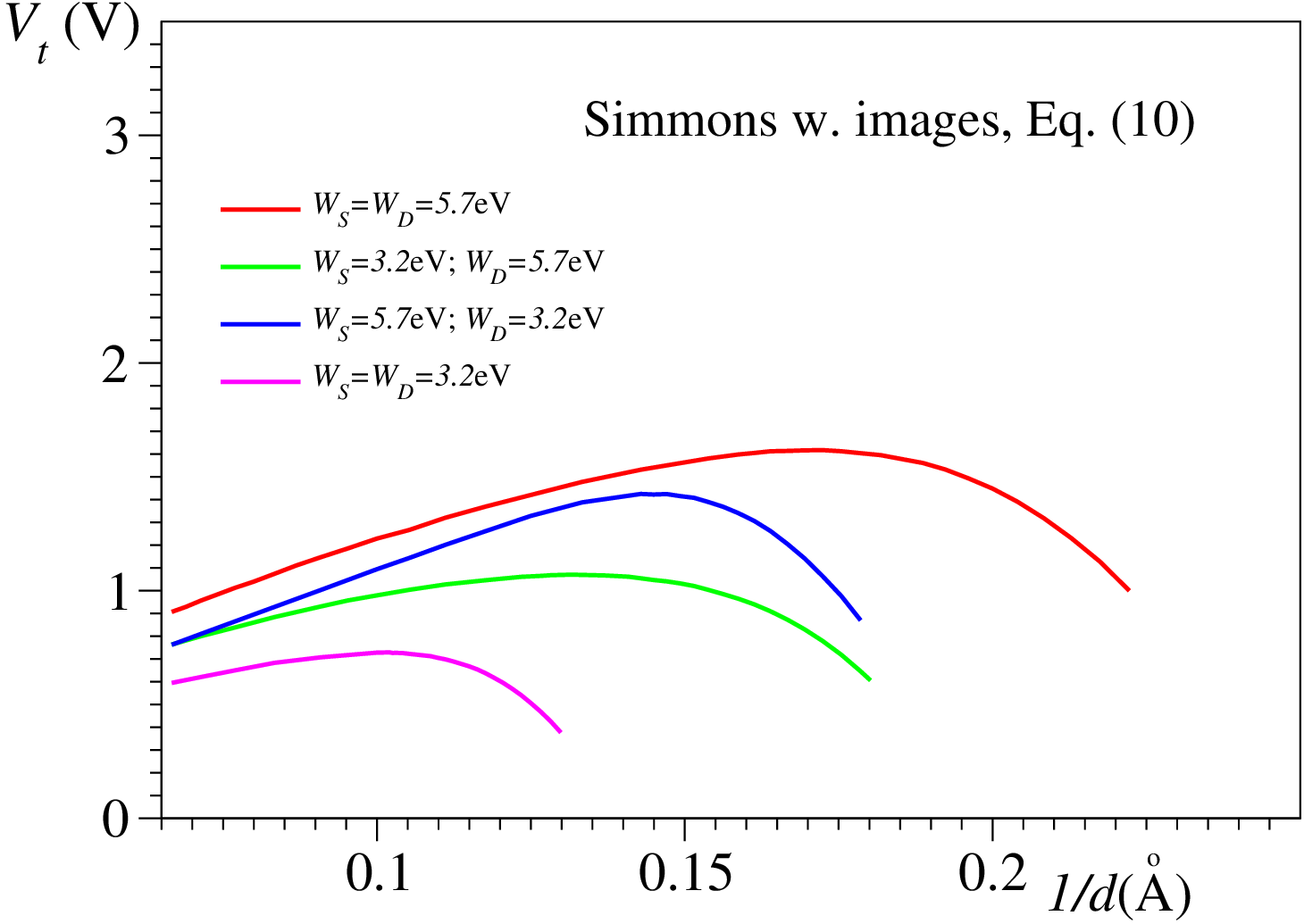}}
$ $\\[6ex]
\centerline{\hspace*{-0ex}\includegraphics[width=0.35\textwidth,angle=0]{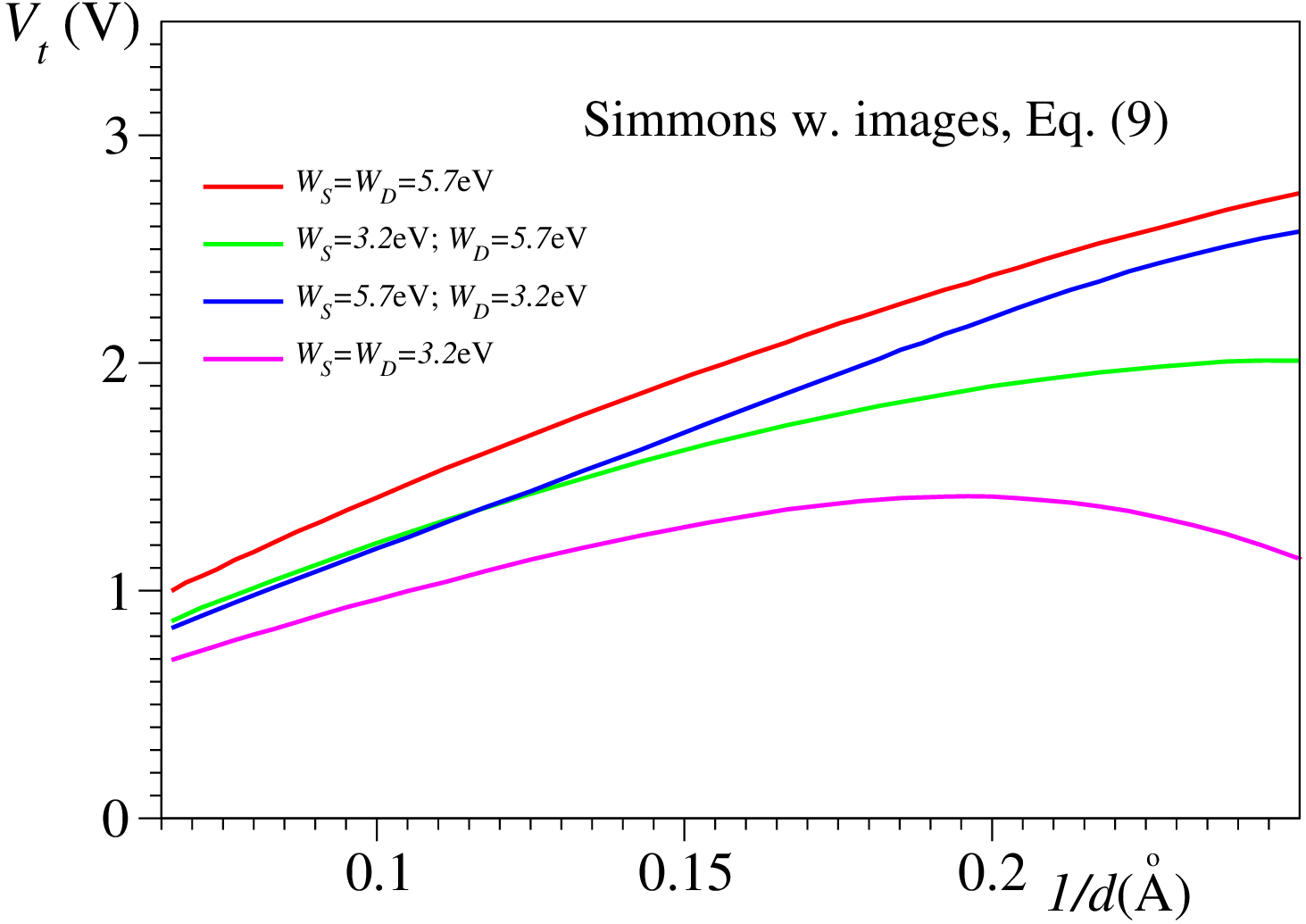}}
\caption{
Transition voltage for vacuum tunneling considering image effects 
computed exactly, and within the WKB and Simmons methods for electrodes with equal and different work functions $W_{S,D}$ specified in the legend.
Simmons curves computed both with the incorrect [Eq.~(\ref{eq-phi-i-simmons-approx}), 
panel $c$] and with the correct [Eq.~(\ref{eq-phi-i}), panel $d$] image forces are shown.
Notice that, with our choice, the source/drain ($S$/$D$) is the negative/positive electrode 
and $V_t$-values are always positive.
}
\label{fig:vt-exact-wkb-huisman-i}
\end{figure}

One can easily convince oneself that the $I$-$V$-characteristics deduced from 
Eqs.~(\ref{eq-phiB}), (\ref{eq-I}) [or (\ref{eq-J-3d})] by using the barrier (\ref{eq-phiB})
are symmetric, $I(V) = -I(-V)$. As a consequence, the transition voltages deduced
for positive and negative biases are of equal magnitude. However,
this symmetry becomes broken in the cases where the work functions are different.
As an illustration of this asymmetry, 
in Figs.~\ref{fig:vt-exact-wkb-huisman-n} and \ref{fig:vt-exact-wkb-huisman-i}, 
we present results obtained by employing 
the experimental values of the work functions $\{W_{S},W_{D}\} = \{5.7, 3.2\}$\,eV
mentioned above, both by considering and ignoring image effects. 
The curves for the transition voltages computed exactly exhibit the aforementioned 
asymmetry both without and with charge images, in qualitative agreement with experiment. 

The work function asymmetry ($W_{S} \neq W_{D}$) 
can represent a general source for asymmetric $I$-$V$-characteristics for 
nanojunctions and even for devices based on symmetric molecules. 
In molecular devices, it can further enhance 
the asymmetry of the electric potential profile obtained by solving the Poisson-Schr\"odinger 
equations selfconsistently.

As in the case of identical electrodes, the Simmons curves for $V_t$ are unacceptable. 
By ignoring image effects, the
Simmons curves are not asymmetric (cf.~Fig.~\ref{fig:vt-exact-wkb-huisman-n}c), 
while by considering image effects all these curves exhibit a maximum
(cf.~Fig.~\ref{fig:vt-exact-wkb-huisman-i}c), which
is absent for the exact curves and is nothing but an artefact of an inadequate approximation.
\section{Apparent barrier height versus work function}
\label{sec:Delta-vs-W}
\begin{figure}[htb]
$ $\\[6ex]
\centerline{\hspace*{-0ex}\includegraphics[width=0.35\textwidth,angle=0]{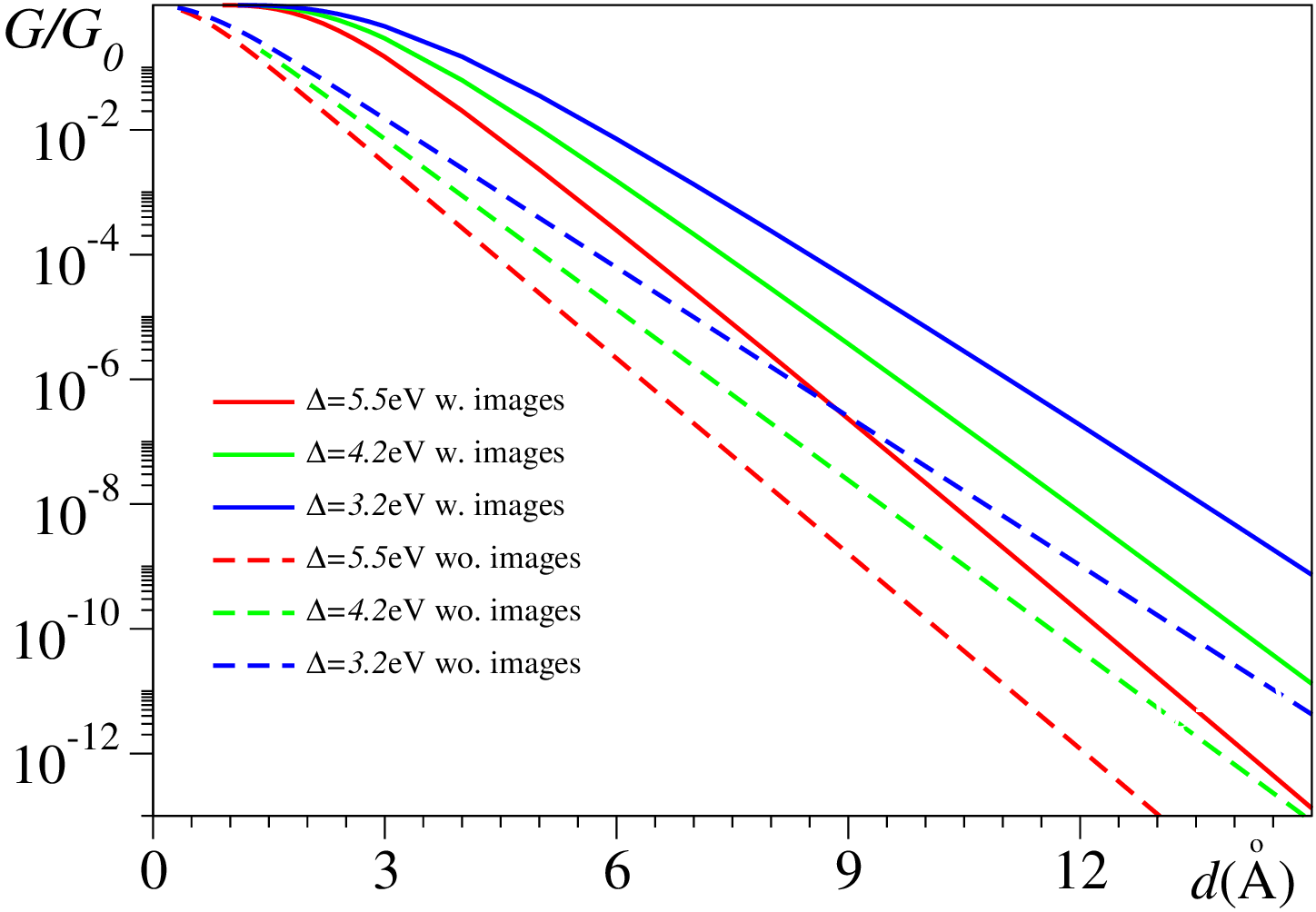}}
\caption{Normalized conductance $G/G_0$ ($G_0 \equiv 2 e^2/h$) computed with and without 
image effects versus nanogap width $d$ for 
several apparent barrier heights $\Delta$ given in the legend. Without images 
$\Delta = W$, while with images $\Delta \simeq W - 0.57$\,eV.}
\label{fig:g-vs-d}
\end{figure}

In the calculations presented above we have employed the (extreme) values of the work functions 
$W_{S,D}$ ($3.2$\,eV and $5.7$\,eV) given in Ref.~\citenum{Molen:11}.
These values have been deduced there from tunneling conductance $G$ 
measurements by assuming a dependence 
$G \propto \exp(-2 d \sqrt{2 m \Delta}/\hbar)$ at larger $d$,\cite{Binnig:84}
and identifying the apparent barrier height $\Delta$ with $W$.
In Fig.~\ref{fig:g-vs-d}, we present results for the conductance computed 
exactly with and without charge images. In the absence of image effects, 
the curves $G(d)$ demonstrate a virtually perfect exponential decay 
down to $d\to 0$ and 
the deduced apparent height is 
$\Delta = \overline{W} \equiv\left(W_{S} + W_{D}\right)/2$.
By including the 
image forces, an exponential decay is still visible at larger $d$. However,
at smaller $d$-values deviations from the exponential decay are clearly visible,
which set in the faster, the lower the work functions are.
The apparent height deduced from the large-$d$ portion 
is no more equal to the average work function,
$\delta \Delta \equiv \Delta - \overline{W} \neq 0$. 
From our exact numerical data, we found a 
difference $\delta \Delta$, which does not depend on $W_{S,D}$ 
and amounts $\delta \Delta \simeq -0.57$\,eV. 
To further check that this difference
is a genuine image effect, we also performed similar calculations by  
artificially increasing the dielectric constant $\kappa_r$ from the vacuum value 
($\kappa_r \equiv 1$). In agreement with the presence of $\kappa_r$ in the denominator 
of the RHS of Eq.~(\ref{eq-phi-i}), we deduced from our numerical results 
a virtually perfect scaling
$\delta \Delta(\kappa_r) \simeq -0.57$\,eV$/\kappa_r$.

Our exact numerical results indicate that even in the presence of image forces, 
the work function asymmetry $\delta W \equiv W_{S} - W_{D} \neq 0$ has practically no effect on 
the linear tunneling conductance;
it  solely depends on the average value $\overline{W}$. 
In contrast to the case at high voltages,
the most exotic asymmetric values of $W_{S,D}$ can hardly change the linear 
conductance by one percent, so we can safely ignore this effect at low voltages $V$.
Still, this behavior indicates that the linear conductance data do not suffice, and 
a further refinement is needed to estimate the values of $W_{S,D}$ from the experimental data. 

Attempting to undertake this effort seems to make little sense at present.
Although relatively large, the above $\delta \Delta$-value is still smaller 
than the experimental inaccuracy of $\sim 1$\,eV of the apparent barrier 
heights deduced in Ref.~\citenum{Molen:11}.
Therefore,
in our numerical calculations, we have simply taken over the experimental values $W_{S,D} = 3.2; 5.7$\,eV 
deduced without applying corrections due to the fact that 
$\delta \Delta \neq 0$ and $\delta W \neq 0$.
\section{On the validity of the WKB method}
\label{sec:wkb}
The results for the transition voltage presented above showed that, 
qualitatively, the ''exact'' WKB curves for $V_t$ behave similar to the exact ones;
quantitatively, differences at smaller sizes are significant. Concerning the
WKB method, we make the following comment at this point. 
The exact counterpart of Eq.~(\ref{eq-wkb-approx}), 
the transmission through a rectangular barrier hill ($\overline{\varepsilon}_B > E_x$)
is well known 
\begin{eqnarray}
\label{eq-met}
\mathcal{T}(E_x, V) & = & 
4 k_{S} k_{D} \kappa_{B}^2\left/ \right.
\left[
\kappa_B^2 \left( k_{S} + k_{D} \right)^2 \cosh^2(\kappa_B s) 
\right. \nonumber \\
& + & \left.
\left( k_{S} k_{D} - \kappa_{B}^2 \right)^2  \sinh^2(\kappa_B s)  \right] ,
\end{eqnarray}
where $\kappa_{B} \equiv [2 m ( \overline{\varepsilon}_B - E_x) ]^{1/2}/\hbar$
and $k_{S,D}$ have been defined above.
[Notice that 
for rectangular barriers, the Simmons and WKB expressions coincide, 
cf.~Eqs.~(\ref{eq-wkb-exact}) and (\ref{eq-wkb-approx}).]
Based on the fact that $\cosh z \approx \sinh z \approx \exp(z)/2$ 
for large $z$,  
it was claimed that the WKB results should not 
notably differ from the exact ones if \cite{Gundlach:69b} 
\begin{equation}
\label{eq-r}
r \equiv A \sqrt{\overline{\varepsilon}_B}/4 \agt 1.
\end{equation}
However, by inspecting the curves for $r$ in Fig.~\ref{fig:vt-gundlach}, one must conclude 
that Eq.~(\ref{eq-r}) does not guarantee reliable WKB and Simmons estimates for $V_t$.
Although the hyperbolic functions entering Eq.~(\ref{eq-met}) reduce to an exponential
for large arguments, there still remains an important difference between the 
exact and WKB transmission, namely the preexponential factor, which is
energy- and $V$-dependent. Were this factor constant, $I_{WKB} = {const} \times I_{exact}$,
the WKB and exact FN-curves $\log (I_{WKB}/V^2) = \log (I_{exact}/V^2) + {const}$ 
would have the minima at the same (transition-)voltage, 
but our results of Figs.~\ref{fig:vt-gundlach}, 
\ref{fig:vt-exact-wkb-huisman-n}, and \ref{fig:vt-exact-wkb-huisman-i}
contradict that $V_t^{WKB}$ represents a reliable estimates for $V_t^{exact}$.
So, the folkloristic dictum on a slowly varying prefactor multiplying a rapidly varying 
exponential does not hold for cases of interest in molecular/nano-transport. 
The need to reconsider the prefactor 
in the FN-tunneling theory for field-emission, a topic of traditional vacuum electronics, 
has also been pointed recently.\cite{Forbes:08} 
As is well known from textbooks, 
the validity of the WKB approximation is difficult to assess in general.
Conclusive evidence only comes from the direct comparison with the exact results,
largely rendering the WKB calculations superfluous.

Although the WKB curves for $V_t$ can significantly deviate from the exact ones,
at least they are qualitatively correct;
they do not exhibit the spurious maximum of the Simmons curves.
Therefore, one may still ask what is the key difference between the 
the WKB and Simmons approximations. In the latter, the actual barrier $\phi_{B}(x)$ 
is replaced 
by a \emph{rectangular} barrier, and this comprises two aspects. 
A rectangular barrier does not only mean a constant height $\overline{\varepsilon}_{B}$
[as if the potential drops at the contacts were the same (namely, $V/2$)], 
but also a constant (energy-independent) barrier width $s = s_2 - s_1$, cf.~Eq.~(\ref{eq-s1s2}).
Let us examine Fig.~\ref{fig:phi-B}. Within the WKB method, depending on their 
energy, the electrons with energies in the Fermi window 
($\mu_{S} = 0 > E_x > \mu_{D} = - e V$) that contribute to the tunneling 
current ``see'' a barrier $\phi_B(x) - E_x$ 
whose width $x_2(E_x) - x_1(E_x)$ is $E_x$-dependent. 
Here, $x_{1,2}$ are defined by $ \phi_B(x_{1,2}) = E_x$. 
Within the Simmons approximation,
electrons tunnel across a rectangular barrier $\overline{\varepsilon}_B - E_x$ 
whose width $\Delta s = s_2 - s_1$ is the smallest;
$x_1(E_x) \leq s_1$ and  
$x_2(E_x) \geq s_2$. Practically, all electrons  
that contribute to the WKB-current tunnel through a barrier broader than the Simmons barrier, 
and this is the most
important reason why the Simmons method 
(even letting alone the missing factor $1/2$) overestimates the image effects.
The approximation $x_2(E_x) - x_1(E_x) \approx s_2 - s_1$ may be justifiable for low biases
($\mu_{S} - \mu_{D} = e V \ll \varepsilon_B$), 
but at higher biases the width of the effective barrier is 
significantly larger than $s_2 - s_1$. As a remedy, one can determine the effective width by using 
the barrier midway between $\mu_{S}$ and $\mu_{D}$, i.~e., using 
$ \phi_B(x)\vert_{x=X_{1,2}} = (\mu_{S} + \mu_{D})/2 = - e V/2 $
instead of Eq.~(\ref{eq-s1s2}). 
This attempt is a trade off of enhanced (reduced) tunnel probabilities 
of electron states below (above) the orange 
line of Fig.~\ref{fig:phi-B}, which wipes out the spurious maximum of the 
Simmons curves discussed above, but it still represents an approximate treatment
more restrictive than the ``exact'' WKB approximation of Eq.~(\ref{eq-wkb-exact}).
\section{Summary and outlook}
\label{sec:conclusion}
The results reported in the present paper can be summarized as follows:

(i) We have drawn attention on the fact that the utilization of the 
barrier picture for molecular or nano-junctions with atomic contacts 
should properly account for the confinement of the transverse electron motion,
a feature not included into the Simmons formula (\ref{eq-j-simmons}).

(ii) We have presented a detailed comparison of the results for the transition
voltage obtained exactly and within the Simmons model.
Particular emphasis has been laid on the fact that the Simmons method,
whose formulas have been utilized to process valuable experimental data 
in several recent studies, is inadequate for molecular/nano-junctions.
Letting alone that the Simmons model (a) ignores the transverse confinement,
(b) it drastically overestimates the image effects because of the missing factor 1/2
in the employed expression of the image interaction energy, and (c) predicts a 
spurious maximum in the curve for $V_t(d)$, which is an artefact of a specific 
rectangular barrier approximation.

(iii) We have presented results demonstrating that, as far as the transition voltage
is concerned, the validity of the WKB method was too optimistically assessed previously.

(iv) We have discussed that a nonvanishing difference bewteen the electrodes' 
work functions, which is likely the case in most experimental setups based on
atomic contacts, may be significant for the ubiquitous asymmetry 
of the $I$-$V$-characteristics 
and the different $V_t$-magnitudes at opposite bias polarities.

Despite intensive and extensive theoretical and experimental efforts 
in the last decade, 
the field of molecular transport is still confronted with a series of difficulties.
The transport problem is alleviated when electrons tunnel through a vacuum nanogap,
and one needs no more to consider,
e.~g., how the positions and widths of the molecular levels 
change upon contacting to electrodes, 
what is the role played by molecular correlations, what is the potential 
profile across the embedded molecule, \emph{etc}, and all these
in a nonequilibrium current-carrying state. 
Therefore, aiming to gain further insight into the transport at nanoscale by
(also) studying vacuum nanojunctions, as recently 
done experimentally,\cite{Molen:11} appears as a noteworthy attempt. 

Although, for reasons like those enumerated above, the transport through vacuum nanogaps
should be easier to understand, it turns out to be far from trivial.
The present study demonstrates that the curves of $V_t = V_t(d)$ reported for the vacuum nanojunctions 
investigated experimentally in Ref.~\citenum{Molen:11} are rather challenging.
In the present paper, we have demonstrated that the initial tentative attempt to 
explain the weak and broad maximum (almost a broad plateau) of the 
experimental $V_t(d)$-curves within 
Simmons-type calculations\cite{Molen:11} (letting alone that the Simmons maxima 
are much more pronounced and narrow than the experimental ones) cannot be justified;
the computed maxima turn out to be an artefact of the Simmons approach.
While being able to indicate the drawbacks of the Simmons approach, we could not
explain the weak $d$-dependence of the experimental curve for $V_t$, although
our results for $V_t(d)$ have been obtained by solving the 
Schr\"odinger equation (\ref{eq-psi-M}) \emph{exactly} and employing the expression (\ref{eq-phi-i}) 
of the charge image contribution, an \emph{exact} result of classical electrostatics.
The weakly $d$-dependent experimental values of $V_t$, which fall in range $\sim 1.5 - 2.5$\,V 
(depending on the apparent barrier height) for $d \alt 8$\,{\AA} \cite{Molen:11}
are significantly smaller than the theoretical estimates of 
Fig.~\ref{fig:vt-exact-wkb-huisman-i}a.
This clearly indicates that other effects, not included within our treatment, should be 
considered, and in the following we will enumerate several possibilities. 

Since the ``exact'' Eq.~(\ref{eq-phi-i}) applies to the case of infinite planar 
electrodes, described as classical continuous media rather than atomic arrays, 
one could attempt to employ a more realistic geometry, or to consider effective  
image planes at $x=x_0$ and $x=d - x_0$ (with $x_0 \sim 2$\,{\AA}), 
displaced from the nominal surfaces at $x=0$ and $x=d$,
as suggested from intuitive reasons \cite{Sommerfeld:33} or by LDA-calculations.\cite{Lang:73} 
The classical result expressed by Eq.~(\ref{eq-phi-i}) can be deduced microscopically as an effect of
surface plasmons in metallic electrodes.\cite{Lenac:76} By considering the static 
limit ($\omega \to 0$) of the surface plasmon polarization, one obtains 
deviations from Eq.~(\ref{eq-phi-i})
\begin{eqnarray*}
\phi_{i}(x) & = &
\frac{e^2}{4 \kappa_r d}
\left[
B_{u}\left(\frac{x}{d}, 0\right) +
B_{u}\left(1 - \frac{x}{d}, 0\right) +
2 \log(1 - u) \right. \\
& - & \left .
\psi\left(\frac{x}{d}\right) -
\psi\left(1 - \frac{x}{d}\right) +
2 \psi(1)
\right]
\end{eqnarray*}
where $u \equiv \exp( - 2 k_c d)$, $k_c$ being a surface
plasmon momentum cutoff, and $B$ is the incomplete 
beta function.\cite{Lenac:76} Eq.~(\ref{eq-phi-i}) can be obtained from the 
above equation by taking the (``local'') limit $k_c \to \infty$.
Retardation effects ($\omega \neq 0$) due to local phonons, surface plasmons \cite{Lenac:76,Tosatti:88}
or finite tunneling time \cite{Weinberg:78} could also play an important role. 
Last but not least, 
in view of the well known fact that
the image potential can bind electrons near the surface of liquid helium,\cite{Grimes:76}
surface image electron states should also deserve consideration. 
A possible role of electron states localized at electrode's surface 
has already been suggested in 
Ref.~\citenum{Molen:11} and discussed more quantitatively in Ref.~\citenum{Lennartz:11}.
\section*{Acknowledgment}
The financial support for this work provided by the Deu\-tsche 
For\-schungs\-ge\-mein\-schaft (DFG) is gratefully acknowledged.

\begin{thebibliography}{10}
\bibitem{why-1.15}
The factor $1.15$ entering Eq.~(\ref{eq-phi-i-simmons-approx}) represents an
  empirical correction, which makes the ``approximate''average value closer to
  the ``exact'' one. Without this factor, the maximum values obtained from
  Eq.~(\ref{eq-phi-i-simmons-ex}) and Eq.~(\ref{eq-phi-i-simmons-approx})
  coincide [$\phi_{i}^{ }(x=d/2) = - 2 e^2 \log 2/(\kappa_r d)$], but near
  electrodes ($x \agt 0$ and $x \alt d$) the ``approximate''values given by the
  latter equation are smaller than the ``exact'' values given by the former
  equation. A significantly better approximation than
  Eq.~(\ref{eq-phi-i-simmons-approx}), which recovers the exact results in the
  limits $x \agt 0$, $x\approx d/2$, and $x \alt d$ is \cite{Lenac:76} $
  \phi_{i}^{ }(x) \approx - \frac{e^2}{4 \kappa_r} \left( 4 \frac{\log 2 -
  1}{d} + \frac{1}{x} + \frac{1}{d - x} \right) $.
\bibitem{integration-limits}
Notice, however, that the Volta field only affects the energy barrier and not
  the integration limits in Eqs.~(\ref{eq-J-3d}) and (\ref{eq-I}).
\end{thebibliography}
\end{document}